# Thermal expansion of $FeWO_4$ (Ferberite) and $FeWO_4$:$Fe_2WO_6$ (7:1): a comparative X-ray and neutron diffraction study

O. Fabelo[1], L. Cañadillas-Delgado[1], D. Vie[2], E. Matesanz[3], J. Gonzalez-Platas[4], and D. Errandonea[5,*]

[1] Institut Laue-Langevin, 71 avenue des Martyrs, CS 20156, 38042 Grenoble cedex 9, France

[2] Institut de Ciència dels Materials de la Universitat de València, Apartado de Correos 2085, E-46071 València, Spain

[3] Unidad de Difracción de Rayos X, Centro de asistencia a la Investigación de Técnicas Químicas, Universidad Complutense de Madrid, Madrid, Spain

[4] Departamento de Física, Instituto Universitario de estudios Avanzados en Física Atómica, Molecular y Fotónica (IUDEA), MALTA Consolider Team, Universidad de La Laguna, La Laguna, Tenerife, Spain

[5] Departamento de Física Aplicada-ICMUV, MALTA Consolider Team, Universitat de Valencia, Dr Moliner 50, Burjassot, Valencia, Spain

*Corresponding author: daniel.errandonea@uv.es

## Abstract

The thermal expansion of natural $FeWO_4$ (ferberite) and synthetic $FeWO_4$:$Fe_2WO_6$ (7:1) was investigated over the 2–1123 K temperature range combining single-crystal and powder X-ray diffraction together with neutron powder diffraction. High-precision lattice parameters were obtained for both samples. The temperature dependence of the unit-cell volume was analysed using physically based thermodynamic models, including the Kroll and Berman approaches as implemented in EoSFit7. All datasets are well reproduced within their respective temperature intervals. However, significant differences are observed between the behavior of ferberite and $FeWO_4$:$Fe_2WO_6$, which has a ~40% smaller thermal expansion coefficient and a reduced reference volume. Possible origins, including microstructural and phase-coexistence effects, are discussed. The results provide a comprehensive description of the thermal expansion behavior of $FeWO_4$ across a wide temperature range.

## 1.- Introduction

Transition-metal tungstates with general formula $MWO_4$ (M = divalent metal) have attracted sustained attention due to their structural versatility and the interplay between lattice, magnetic, and electronic degrees of freedom [1]. Among them, $FeWO_4$ (ferberite) crystallizes in the monoclinic wolframite-type structure (space group *P*2/*c*), characterized by chains of edge-sharing $FeO_6$ octahedra linked through $WO_6$ octahedra (see Figure 1 in the Supporting Information). The crystal structure of ferberite was established in early crystallographic studies [2], and more recently refined and analyzed using modern diffraction techniques [3].

$FeWO_4$ exhibits anisotropic physical properties and long-range antiferromagnetic order below its Néel temperature [4, 5]. Detailed investigations of its magnetic and transport properties have confirmed the coexistence of magnetic ordering and structural stability at low temperatures [6]. The coupling between magnetic ordering and lattice degrees of freedom makes $FeWO_4$ a suitable model compound for investigating thermoelastic behavior in magnetically ordered oxides.

Accurate knowledge of thermal expansion is essential for understanding lattice dynamics, thermodynamic stability, and magnetoelastic coupling in crystalline solids. The temperature dependence of the unit-cell volume provides direct access to fundamental thermodynamic parameters such as the thermal expansion coefficient and characteristic temperatures entering physically based models of lattice expansion. Several approaches have been developed to describe the temperature evolution of volume, including the semi-empirical formalism of Kroll *et al.* [7], the thermodynamic treatment of Salje *et al.* [8], and the high-temperature formulation introduced by Berman [9]. These models are widely implemented in equation-of-state analyses and mineral thermodynamic databases.

Despite the structural and magnetic characterization available in the literature [5, 6, 10], a comprehensive description of the thermal expansion of $FeWO_4$ over a broad temperature interval remains limited. Previous investigations have primarily focused either on magnetic behavior at low temperature [5] or on structural refinements at ambient conditions. Systematic comparisons between different diffraction techniques, particularly single-crystal X-ray diffraction and neutron powder diffraction, are scarce, although such comparisons are relevant because the two techniques differ in scattering contrast, sensitivity to light elements, and response to microstructural effects, which may influence refined lattice parameters and derived thermoelastic quantities.

In this work, we present a detailed investigation of the thermal expansion of ferberite (natural $FeWO_4$) over the temperature range 2–1123 K. We also report results for a synthetic 7:1 $FeWO_4$:$Fe_2WO_6$ composite (two-phase mixture). For clarity, throughout the manuscript the natural mineral is referred to as ferberite, whereas the laboratory-synthesized $FeWO_4$:$Fe_2WO_6$ material is referred to as synthetic $FeWO_4$. High-precision lattice parameters were obtained using single-crystal X-ray diffraction on a natural ferberite sample, complemented by powder X-ray diffraction and neutron powder diffraction measurements on synthetic $FeWO_4$. The temperature dependence of the unit-cell volume was analyzed using physically based thermodynamic models, including the Kroll and Berman approaches as implemented in EoSFit7 [11]. Attention is devoted to the comparison between the two samples, which reveal significant differences in the derived thermal expansion parameters. The possible origins of these discrepancies, including compositional, microstructural, and phase-coexistence effects, are critically evaluated.

By combining complementary diffraction techniques over an exceptionally wide temperature range, this study provides a comprehensive thermoelastic description of ferberite and synthetic $FeWO_4$ and contributes to the understanding of how experimental methodology and sample characteristics may affect the determination of thermal expansion parameters in complex transition-metal oxides.

# 2.- Experimental details

*2.1 Samples*

The experiments were carried out using two types of samples. The first sample was polycrystalline and was synthesized using a co-precipitation following the procedure described by Fabelo *et al.* [10]. It should be noted that the final product consisted of a two-phase mixture of $FeWO_4$ (ca. 88%) and $Fe_2WO_6$ (ca. 12%). The second sample was of natural origin (ferberite) obtained from the Monte Cambillaya mining district, La Paz, Bolivia, and was used in previous studies [3]. Nb (0.06%) and Ta (0.02%) were the only detected impurities, thus the sample can be considered phase pure. The natural material consisted of large single crystals, which were subsequently ground to prepare polycrystalline samples.

*2.2 X-ray single-crystal diffraction*

Temperature-dependent single-crystal diffraction (SC-XRD) data on ferberite were collected on the dual-source Bruker D8 diffractometer at the Institut Laue–Langevin (ILL). The mounted crystal used for the SC-XRD experiment is shown in Figure 1. Silver radiation with a wavelength of λ = 0.56086 Å was used for all measurements. Six ω/ϕ scans were collected using different counting times to achieve a resolution of 0.5 Å. The collected frames were integrated using the Bruker SAINT software package [12] with the narrow-frame algorithm. Data integration was performed using a monoclinic unit cell, which was refined at each temperature prior to integration. The final unit cell at each temperature was determined using the XYZ-centroids method, considering only reflections with $I > 20\sigma(I)$ and without applying any angular cut-off.

Two different temperature setups were employed: a Helix Cryostream for measurements between 40 and 300 K, and a Cobra Cryostream for measurements between 300 and 500 K. The change of setup was performed without modifying the crystal orientation, to avoid any significant change in the orientation matrix that could induce artefacts in the data analysis.

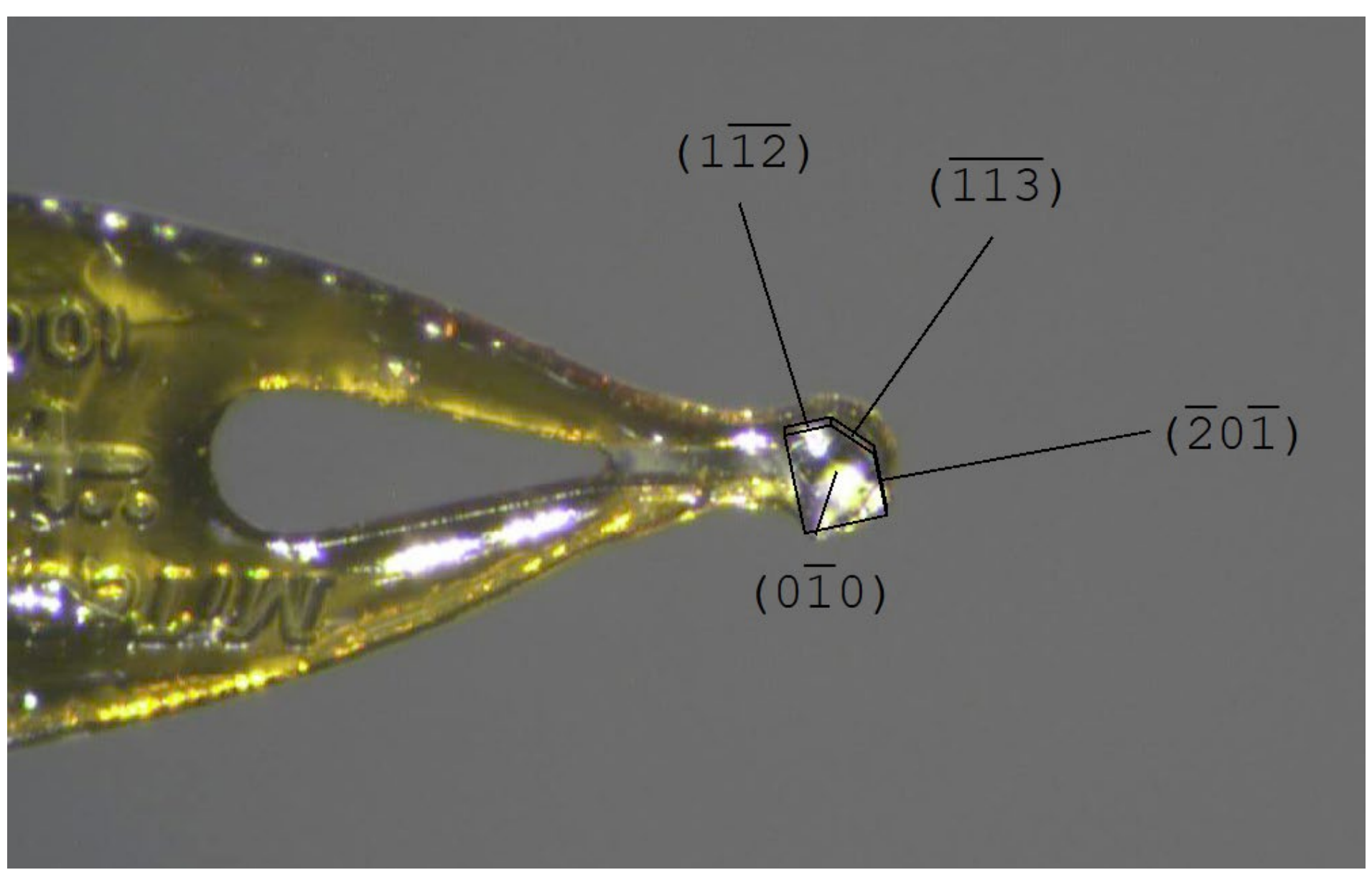

**Figure 1.** View of the single crystal of ferberite used in the SC-XRD experiment in the Bruker D8 diffractometer.

*2.3 X-ray powder diffraction*

X-ray diffraction measurements on the powdered ferberite sample and synthetic $FeWO_4$ were carried out using an X'Pert diffractometer equipped with Cu $K_\alpha$ radiation and an X'Celerator detector at Universidad Complutense de Madrid. Low-temperature measurements in the 15–300 K range were performed using a Phenix low-temperature chamber (Oxford Cryosystems), featuring a closed-cycle design optimized for this type of diffractometer. High-temperature experiments (298–1123 K) were conducted using an HTK-2000 chamber from Anton Paar. In both cases, measurements were performed under vacuum and in Bragg–Brentano (reflection) geometry. Detailed results from experiments are given in the Supporting Information.

Temperature evolution of the X-ray diffraction patterns were analyzed by Le Bail profile refinements [13] considering two crystalline phases in the synthesized sample, namely $FeWO_4$ (space group *P2/c*) and a minor $Fe_2WO_6$ impurity phase (space group *Pbcn*) and only $FeWO_4$ in ferberite. The refinements were carried out using the FullProf Suite package [14]. The background was modelled using interpolated background points and selected angular regions were excluded from the refinement. The peak profiles were described using a pseudo-Voigt function. In the final cycles, 27 parameters were refined including the zero shift, scale factors of both phases, lattice parameters and profile parameters (U, V, W, and X). Detailed results from experiments are given in the Supporting Information.

*2.4 Neutron powder crystal diffraction*

Neutron powder diffraction data were collected on the high-counting-efficiency D20 diffractometer, operated in high-resolution mode at the Institut Laue-Langevin. The synthesized powder sample was loaded into a cylindrical vanadium can and the measurements were performed as a function of temperature using a standard ILL cryofurnace, allowing temperature control over the 2-500 K range. The temperature was stabilized at each set point prior to data acquisition to ensure thermal equilibrium. The instrument was equipped with a radial oscillating collimator in order to reduce air and sample-environment scattering. Diffraction patterns were recorded over the 2θ range 0°-152.9° with a step size of Δ2θ = 0.1°, using the position-sensitive detector (PSD) based on 3He technology, which provides high counting efficiency and excellent angular stability. Detector efficiency correction was applied to the raw data before the data analysis.

The incident neutrons were monochromated using the (115) Bragg reflection of a germanium (Ge) monochromator with a take-off angle of 90°, corresponding to a nominal wavelength of 1.54 Å [15]. The exact wavelength (λ = 1.54119(3) Å) was determined by refinement of a standard $Na_2Ca_3Al_2F_{14}$ sample at room temperature, constraining the lattice parameters to values previously obtained under the same conditions from high-resolution X-ray powder diffraction data.

Neutron diffraction patterns were analyzed by Rietveld refinements considering both nuclear and magnetic contributions. Below the Néel temperature, the nuclear and magnetic structures of synthetic $FeWO_4$ were included in the refinement together with the nuclear contribution of a minor $Fe_2WO_6$ impurity phase. The refinements were carried out using the FullProf Suite package [14]. The instrumental resolution function was determined from a standard reference sample measured under identical instrumental conditions. The background was modelled using interpolated background points and selected angular regions were excluded from the refinement. The peak profiles were described using a Thompson-Cox-Hastings pseudo-Voigt function convoluted with an axial divergence asymmetry function following Finger, Cox, and Jephcoat [16]. Data visualization and export were performed using the PANDA program [17]. Detailed results from experiments are given in the Supporting Information.

## 3.- Results and discussion

The volume thermal expansion of a material can be expressed as

$$V(T) = V_0\, exp \int_{T_{ref}}^{T} \alpha(T) dT$$

where $T_{ref}$ is a reference temperature at which volume is $V_0$. The only thermodynamic constraints on the form of the function for $\alpha(T)$ are that both $\alpha(0)$ and $\partial\alpha(T)/\partial T$ vanish at absolute zero. Additionally, it is observed in many experiments that at high temperatures $\alpha(T)$ increases linearly with temperature.

Despite the large number of models proposed in the literature, not all of them fully satisfy these requirements. Some models can account for the low-temperature saturation but, conversely, fail to provide physically reasonable fits at high temperatures, whereas other models exhibit the opposite behavior. Nevertheless, many of these models are regarded as valid within the temperature ranges, they are intended to describe and are therefore widely used in thermodynamic databases.

Given that this compound exhibits long range magnetic order at low temperatures, an accurate description of its thermal expansion in the low temperature regime is required. To model the experimental *V(T)* data, we employed the approach proposed by Kroll *et al.* [7] which explicitly relates the volume to the lattice energy of the material, as well as the model suggested by Salje *et al.* [8]. Both approaches are implemented in the EoSFit7 software package [11]. The results obtained from fitting the experimental data are indistinguishable; therefore, only the calculations based on the Kroll model are presented. In this model, the parameters $V_0$, $\alpha_0$, and $\theta_E$ were refined. Here, $\alpha_0$ is the thermal expansion coefficient at $T_{ref}$ and $\theta_E$ is the Einstein temperature, which is related to the molar standard-state entropy and accounts for the saturation of $\alpha(T)$ at low temperatures. The parameter $K'$, corresponding to the first derivative of the bulk modulus at $T_{ref}$, was fixed according to the experimental results reported in high-pressure XRD experiments [3].

To obtain high-precision thermal expansion data, we first performed single-crystal diffraction measurements using the natural ferberite crystals previously studied in Ref. [3]. To exclude possible effects arising from the single-crystal indexing procedure, part of the ferberite sample was ground and sieved to prepare a high-quality powder specimen. This powder was subsequently used for temperature-dependent X-ray diffraction measurements in the polycrystalline form. In both cases the results obtained are comparable, taking as a reference the value at a temperature of 298.0(5) K. The results obtained for the temperature-dependence of the volume are represented in Figure 2. Both experiments give similar behavior. The unit-cell volume obtained from powder XRD differs slightly from that determined by SC-XRD (≈0.2%). Such a small discrepancy is within the typical uncertainty expected from measurements performed using different experimental setups. The parameters obtained from the description of the results using the Kroll model are summarized in Table 1. The values determined for $\alpha_0$ from both experiments agree within uncertainties.

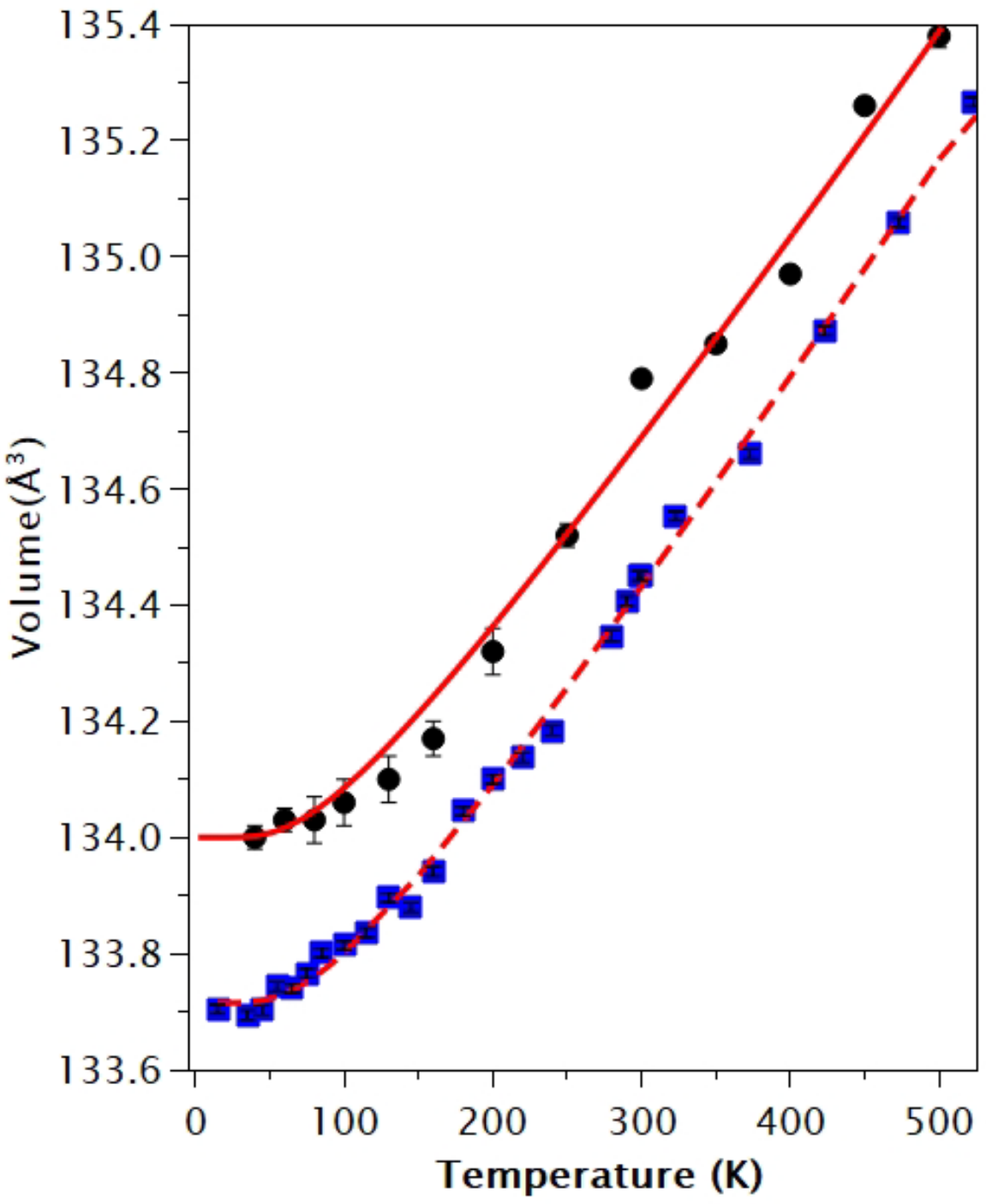

**Figure 2.** Temperature evolution of the unit-cell volume for the ferberite sample determined by X-ray diffraction. Single-crystal measurements (circles) were collected between 40 and 500 K, while powder X-ray diffraction measurements on the corresponding ground sample (squares) were performed between 15 and 523 K. Symbols represent the experimental values with their corresponding error bars. Solid lines correspond to the fit of the data using the Kroll model.

| | Ferberite | | $FeWO_4$:$Fe_2WO_6$ | |
|---|---|---|---|---|
| **Kroll model** | **SC-XRD** | **XRD** | **Neutron** | **XRD** |
| $V_0$ (Å$^3$) | 134.683(15) | 134.425(7) | 132.392(4) | 132.148(8) |
| $K'^*$ | 5.0 | 5.0 | 5.0 | 5.0 |
| $\alpha_0(\times 10^{-5}K^{-1})$ | 2.50(11) | 2.63(4) | 1.79(4) | 1.67(6) |
| $\theta_E(K)$ | 235(57) | 242(18) | 256(15) | 389(37) |

**Table 1.** Parameters obtained from fitting the data using the Kroll model. Parameters marked with an asterisk (*) were fixed during the refinement to improve the stability of the fit and avoid overparameterization. In particular, the pressure derivative was $K'_0$ fixed to the value reported in the literature [3].

To accurately characterize the low-temperature structural evolution and to determine the temperature dependence of both the nuclear and magnetic structures, we performed neutron powder diffraction measurements on the D20 instrument at ILL. In this work we will focus on the behavior of the crystal structure, leaving the analysis of the nuclear and magnetic structures to a future study. Due to the large amount of material required for neutron experiments, we discarded the use of the natural sample and instead employed a freshly synthesized polycrystalline specimen. As described above, in synthetic $FeWO_4$, two phases were detected in an approximate 7:1 ratio of $FeWO_4$ and $Fe_2WO_6$. The degree of overlap between the two crystalline phases is not high and allows us to evaluate the behavior of $FeWO_4$ as a function of temperature without apparent problems (see Figure 3).

Neutron diffraction data further demonstrate that no significant anomalies are observed in either the unit-cell volume or the lattice parameters as a function of temperature. This absence of discontinuities confirms that no additional structural contributions interfere with the present analysis. The magnetic structure below $T_n$ (ca. 62K) is well described by the AF1 collinear configuration [10], with no evidence of additional magnetic phases or concomitant structural distortions. The refined magnetic moments and their temperature dependence do not reveal any anomalous behavior indicative of strong magnetoelastic coupling. In particular, the absence of measurable changes in interatomic distances or symmetry across $T_n$ indicates that magnetostrictive effect is weak.

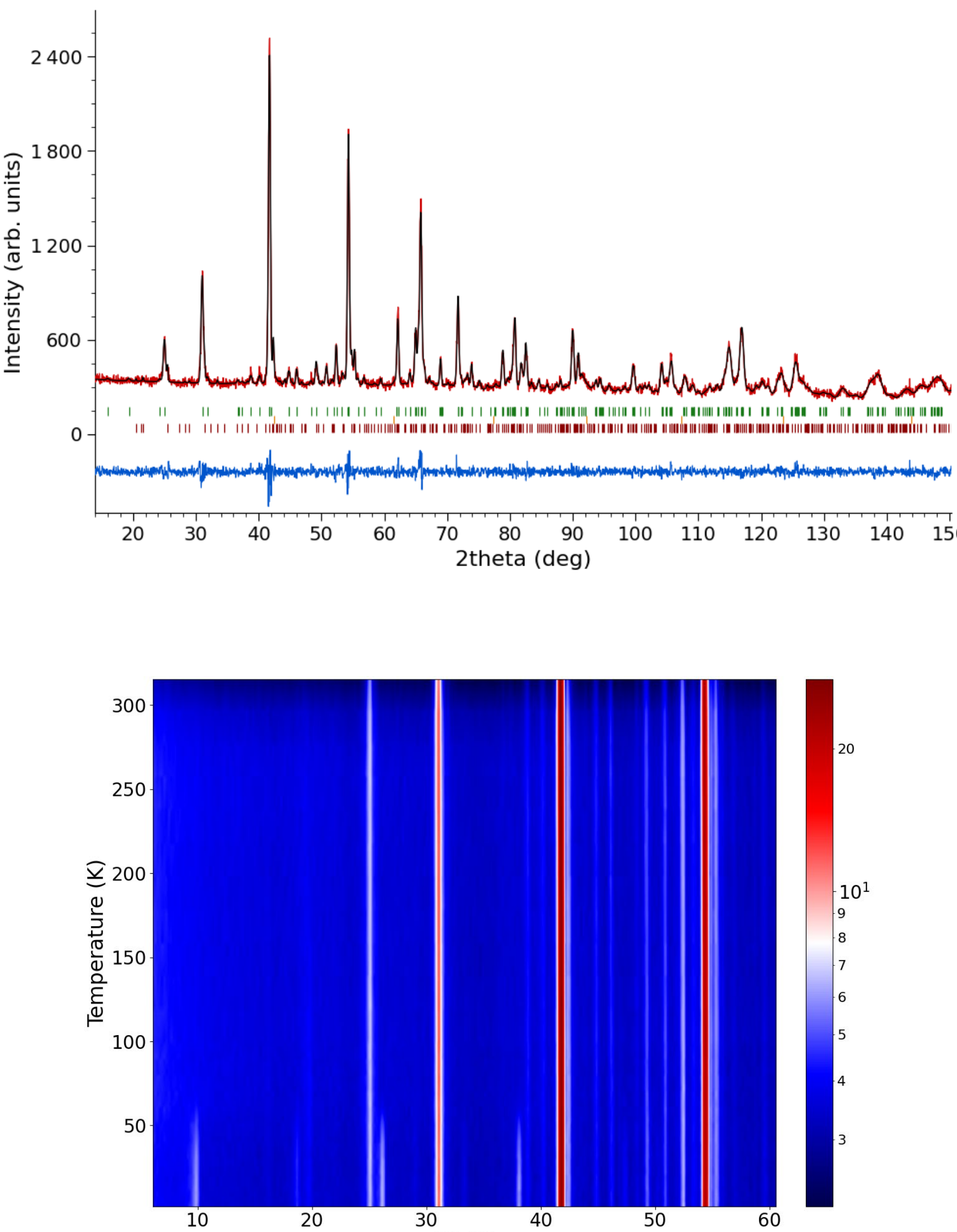

**Figure 3.** (Top) Rietveld refinement of the neutron powder diffraction pattern of synthetic $FeWO_4$ measured at D20 at 300 K (λ = 1.54119(3) Å). The experimental data and calculated profile are shown as red and black lines, respectively, while the blue curve represents the difference. Vertical green, orange, and red ticks indicate the Bragg reflection positions of the different phases. The second phase corresponds to V from the sample holder, whereas the third phase is attributed to $Fe_2WO_6$, refined in the orthorhombic *Pbcn* space group with lattice parameters $a$ = 4.6000(6), $b$ = 16.814(3), and $c$ = 4.9674(7) Å. This contribution was modeled using a Le Bail approach, leading to a final $\chi^2$ = 1.26. (Bottom) two-dimensional thermodiffractogram obtained from neutron powder diffraction measurements of synthetic $FeWO_4$ at D20, plotted on a logarithmic intensity scale. Magnetic Bragg reflections associated with the propagation vector **k** = (1/2, 0, 0) appear below approximately 62 K.

In Figure 4 we present the results obtained from synthetic $FeWO_4$. From both experiments we determined a similar temperature dependence. The parameters obtained from the description of the results with the Kroll model are given in Table 1. The values of the volumes obtained from both methods differ by less than 0.2%. Comparing the results obtained from ferberite and synthetic $FeWO_4$, we observe a discrepancy in the refined thermal expansion parameters. In particular, the thermal expansion coefficient $\alpha_0$ derived from synthetic $FeWO_4$ (from both neutron and XRD diffraction) differs by more than 40 % from the value obtained from ferberite (from powder XRD and SC-XRD). In addition, the unit-cell volume at the reference temperature is systematically smaller in synthetic $FeWO_4$ than in ferberite. This result is unexpected, as both experiments probe the same crystallographic phase, and its origin will be discussed in detail below.

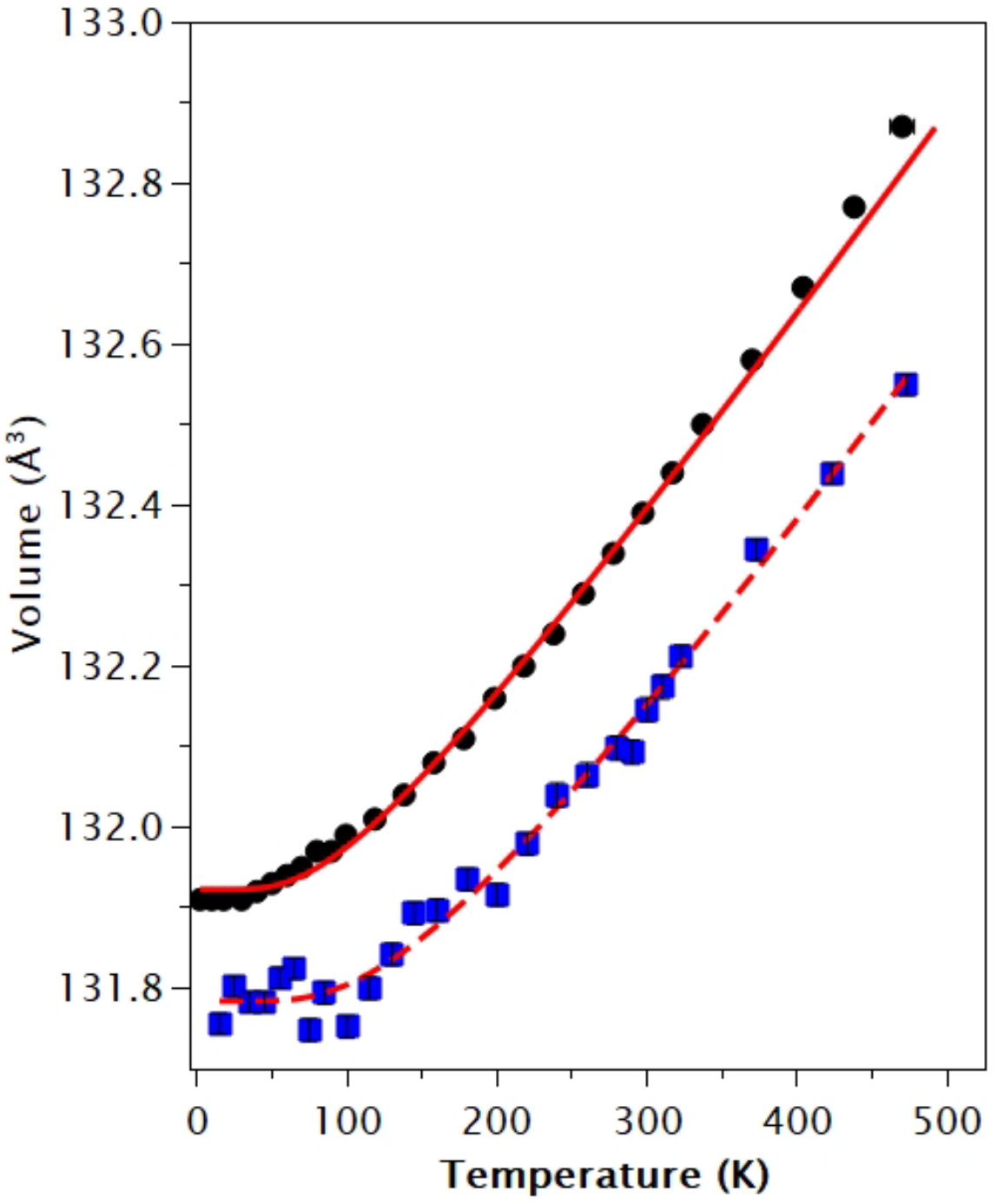

**Figure 4.** Results from the experiments performed in synthetic $FeWO_4$. Volume behavior between 2 and 470 K measured using neutrons in D20 diffractometer at ILL (circles) and 15-473 K results obtained from powder XRD (squares). Error bars are smaller than symbols. The lines show the fit of the data to Kroll model.

A reduced unit-cell volume in the synthesized material could in principle indicate either the presence of larger-radius ions incorporated in the lattice of the natural $FeWO_4$ sample, or, conversely, partial substitution by smaller-radius ions in the synthesized compound. Both scenarios can be reasonably excluded. The incorporation of heavier or larger-radius transition-metal or rare-earth impurities in the natural sample was ruled out based on EDX measurements in previous studies [2], which indicates negligible concentrations of contaminant elements. Likewise, substitution effects in the synthesized sample are unlikely, since the chemical composition is controlled by the starting reagents used during synthesis.

Experimental artefacts related to temperature calibration, sample centering, or uncertainties in the incident wavelength were also discarded, as the powder XRD in ferberite and synthetic $FeWO_4$ were performed under identical experimental conditions. Finally, although the measurements in synthetic $FeWO_4$ were carried out on a two-phase polycrystalline sample ($FeWO_4$:$Fe_2WO_6$ ≈ 7:1), peak overlap between both phases is limited and does not prevent reliable refinement of the $FeWO_4$ lattice parameters, moreover both phases were included in the LeBail refinements for all temperatures. On the other hand, microstructural effects such as strain or intergrowth defects typically lead to an apparent lattice expansion rather than a contraction [18] and therefore do not provide a straightforward explanation for the smaller refined volume observed in synthetic $FeWO_4$.

We consider that the smaller volume of synthetic $FeWO_4$ compared to ferberite might be related to a structural distortion related to lattice strain. If $Fe_2WO_6$ forms at grain boundaries or partially incorporates into $FeWO_4$, it can induce compressive lattice strain inducing a lattice contraction of 2% in synthetic $FeWO_4$ [19]. Another hypothesis that might explain the contraction of the volume in synthetic $FeWO_4$ might be related to a partial oxidation of $Fe^{2+}$ → $Fe^{3+}$ or to the presence of oxygen vacancies. Since $Fe^{3+}$ has a smaller ionic radius than $Fe^{2+}$, this can shrink lattice parameters. This is very common in tungstates [1].

Regarding the difference in thermal expansion, it may originate from the presence of the $Fe_2WO_6$ secondary phase in Synthetic $FeWO_4$ acting as a mechanically constraining component. In two-phase systems, the effective coefficient of thermal expansion (CTE) depends not only on the intrinsic CTEs of the individual phases but also on their elastic interaction; a phase with a lower thermal expansion coefficient and/or higher bulk modulus can restrict the expansion of the matrix phase, leading to a reduced overall CTE of the composite [20]. However, our temperature-dependent single-crystal measurements do not reveal any significant variation in the Fe–O bond distances. The presence of $Fe^{3+}$ would be expected to induce a measurable contraction of these distances; however, no such effect is observed. Instead, the data show only a slight increase in the Fe–O2 bond length (see Table S2 in the Supporting Information), which is consistent with a moderate anisotropic thermal expansion.

This absence of bond contraction therefore argues against a significant contribution of $Fe^{3+}$ to the structure. Similarly, the presence of oxygen vacancies would likely lead to a more heterogeneous distortion of the $FeO_6$ octahedra or to enhanced local disorder, which is not supported by the present structural data. Altogether, these observations suggest that the reduced unit-cell volume in synthetic $FeWO_4$ is more plausibly related to microstructural effects, rather than to intrinsic changes in iron oxidation state.

To our knowledge, no peer-reviewed study has directly reported the thermal expansion coefficient or bulk modulus of $Fe_2WO_6$. However, insight can be gained from related iron oxides. Experimental and computational studies on FeO and $Fe_2O_3$ indicate that $Fe^{3+}$,rich compounds generally exhibit lower thermal expansion coefficients and higher bulk moduli compared with $Fe^{2+}$, dominated systems [21, 22]. By analogy, it is reasonable to expect that $Fe_2WO_6$, which contains a higher proportion of $Fe^{3+}$ relative to $FeWO_4$, would possess a lower CTE and greater stiffness.

This trend is further supported within the Fe–W–O system: $Fe_2W_3O_{12}$ exhibits a smaller thermal expansion coefficient ($1.35 \times 10^{-6}$ $K^{-1}$) than $FeWO_4$ [23]. Notably, the $Fe^{3+}$ content increases along the sequence $FeWO_4$ (FeO + $WO_3$) < $Fe_2W_3O_{12}$ ($Fe_2O_3$ + $3WO_3$) < $Fe_2WO_6$ ($Fe_2O_3$ + $WO_3$). Since $Fe^{3+}$ has a significantly smaller ionic radius than $Fe^{2+}$ [24], increasing $Fe^{3+}$ concentration is expected to strengthen bonding and enhance lattice rigidity, consistent with general correlations between cation size, compressibility, and structural stiffness in transition-metal oxides [25].

In summary, although direct thermoelastic data for $Fe_2WO_6$ are not yet available, chemical and structural considerations strongly suggest that $Fe_2WO_6$ may be mechanically stiffer and thermally less expansive than $FeWO_4$ due to the presence of $Fe^{3+}$. Therefore, its presence in the composite can reasonably be expected to act as a mechanical constraint phase, limiting the lattice expansion of $FeWO_4$; a well-established thermos-mechanical effect in multiphase materials [20].

To further clarify the lattice response, the temperature dependence of the individual lattice parameters has been analyzed (see Figure 5). These data show that both compounds exhibit anisotropic lattice expansion along the three crystallographic directions, with a more pronounced variation along the *c*-axis. This behavior is consistent with the low-symmetry nature of the structure and the anisotropic bonding environments. Interestingly, the natural sample exhibits a more pronounced anisotropy of the lattice parameters, which may be associated with the presence of defects or trace dopant inclusions, even in very small fractions, affecting the directional lattice response. Nevertheless, both compounds display a very similar overall trend, with the degree of anisotropy remaining moderate, and no additional structural anomalies observed in the low-temperature region beyond the expected reduction of thermal expansion upon cooling.

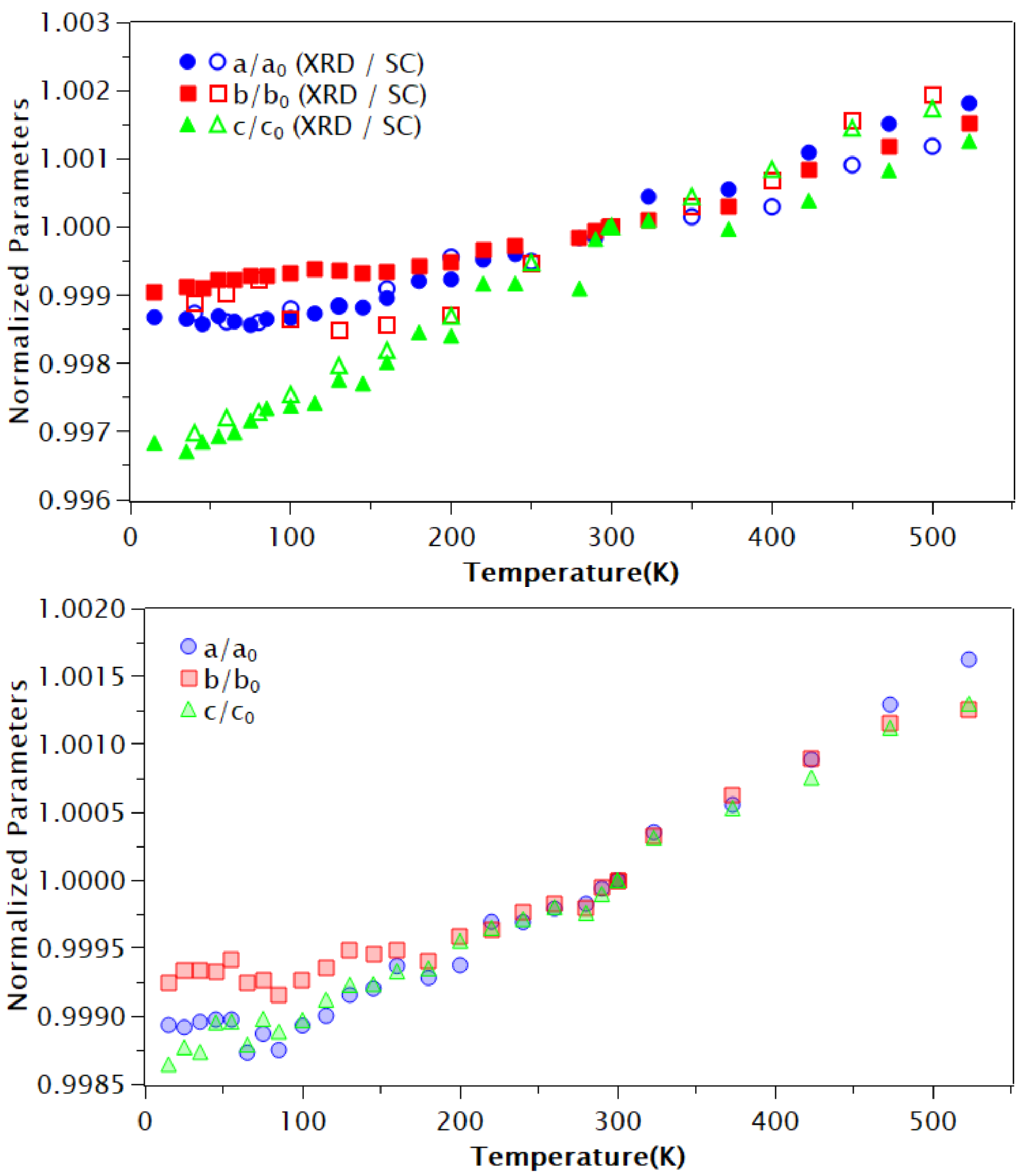

**Figure 5:** (Top) Temperature evolution of the lattice parameters of ferberite derived from single-crystal and powder diffraction measurements (open and filled symbols, respectively). (Bottom) Temperature evolution of the lattice parameters of synthesized $FeWO_4$ obtained from powder diffraction data.

Finally, since the stability of this type of compound is high with temperature, we have completed the study of the thermal expansion of the ferberite sample and synthetic $FeWO_4$ up to 1123 K. Not all existing models accurately describe the thermal expansion behavior of solids. Therefore, in this temperature range (from RT to high temperatures), the model proposed by Berman [9] appears to be the most appropriate. Berman proposed a simple extension to accommodate non-linear thermal expansion

$$V(T) = V_0\left(1 + \alpha_0[T - T_{ref}] + \frac{1}{2}\alpha_1[T - T_{ref}]^2\right)$$

Having small changes in volume with temperature, the thermal expansion coefficient can be expressed as

$$\alpha(T) \approx \alpha_0 + \alpha_1(T - T_{ref})$$

The results of the two experiments are shown in Figure 6. At low temperature the thermal expansion of the synthetic sample is smaller than in ferberite. This is consistent with the results discussed above. The fits to the data using the Berman model are included in the figure. The fitted parameters are shown in Table 2. The results confirm that synthetic $FeWO_4$ has a smaller volume than ferberite. The values obtained for $\alpha_0$ from the high-temperature studies agree with the values obtained in the range of 15 to 523 K using the Kroll model (see Table 1).

| **Berman model** | **Ferberite** | **$FeWO_4$:$Fe_2WO_6$** |
|---|---|---|
| $V_0$ (Å$^3$) | 134.310(14) | 131.97(2) |
| $\alpha_0(\times 10^{-5} K^{-1})$ | 2.75(6) | 1.08(10) |
| $\alpha_1(\times 10^{-8} K^{-2})$ | -1.19(13) | 2.8(3) |

**Table 2.** Parameters obtained from fitting the high-temperature data using the Berman model.

Interestingly, above 800 K, the temperature dependence of the unit-cell volume shows opposite trends in the two samples: a decrease in slope for ferberite and an increase for synthetic $FeWO_4$. As a result, at the highest temperature studied, the thermal expansion of the synthetic sample exceeds that of ferberite. This behavior is reflected in the thermal expansion coefficients, with $\alpha_1$ being negative for ferberite but positive for synthetic $FeWO_4$ (see Table 3). For most solids, $\alpha_1$ is positive, but several factors can cause it to decrease or even become negative at higher temperatures [26]. The negative value of $\alpha_1$ in ferberite may reflect anharmonic saturation effects, microstructural relaxation, or polyhedral rotations, which can hinder the expansion of individual bonds at high temperature [27].

An alternative explanation could involve a partial oxidation of $Fe^{2+}$ to $Fe^{3+}$, which is generally more likely at elevated temperatures. However, no clear experimental evidence supports this scenario in the present case. In agreement with the single-crystal results discussed above, no significant changes are observed in the lattice parameters before and after thermal cycling, nor in the Fe–O bond distances, which do not show the expected contraction associated with $Fe^{3+}$. It should be noted, however, that the determination of oxygen atomic positions from X-ray powder diffraction data in the presence of heavy elements such as W remains challenging, and therefore the precision of the Fe–O distances is limited (see Figure 7).

| | Before heating | | After heating | |
|---|---|---|---|---|
| | Ferberite | Synthetic $FeWO_4$ | Ferberite | Synthetic $FeWO_4$ |
| $a$ /Å | 4.7375(1) | 4.6907(1) | 4.7375(1) | 4.7020(2) |
| $b$ /Å | 5.7096(1) | 5.6953(1) | 5.7096(1) | 5.6946(2) |
| $c$ /Å | 4.9650(1) | 4.9478(1) | 4.9651(1) | 4.95424(1) |
| α /° | 89.899(4) | 89.918(3) | 89.899(3) | 89.837(2) |
| $R_f(\chi^2)$ | 6.68(2.7) | 3.37(1.7) | 6.69(2.5) | 4.14(2.1) |
| | **Ferberite** | | | |
| | Bond distance (Å) | | Bond distance (Å) | |
| $Fe-O1$ | 1.89(3) | | 1.89(3) | |
| $Fe-O2$ | 1.99(2) | | 1.99(2) | |
| $Fe-O2a$ | 2.27(2) | | 2.27(2) | |
| | **Synthetic $FeWO_4$** | | | |
| $Fe-O1$ | 1.96(3) | | 1.94(2) | |
| $Fe-O2$ | 2.08(3) | | 2.06(3) | |
| $Fe-O2a$ | 2.12(2) | | 2.09(2) | |

**Table 3.** Unit-cell parameters, refinement reliability factors, and selected Fe–O distances for ferberite and synthetic $FeWO_4$ compounds, obtained by the Rietveld refinement method using the powder X-ray diffraction data collected in the X'Pert diffractometer, Cu $K_\alpha$ radiation, before and after heating to 1123 K. symmetry operator: a = *x*, -*y*, *z*+1/2.

Instead, the different behavior of synthetic $FeWO_4$ compared to ferberite is more plausibly explained by microstructural effects related to the multiphase nature of the sample. In particular, the presence of $Fe_2WO_6$, which exhibits a distinct thermal expansion, may exert mechanical constraints on the $FeWO_4$ phase, leading to the observed deviation in the temperature dependence.

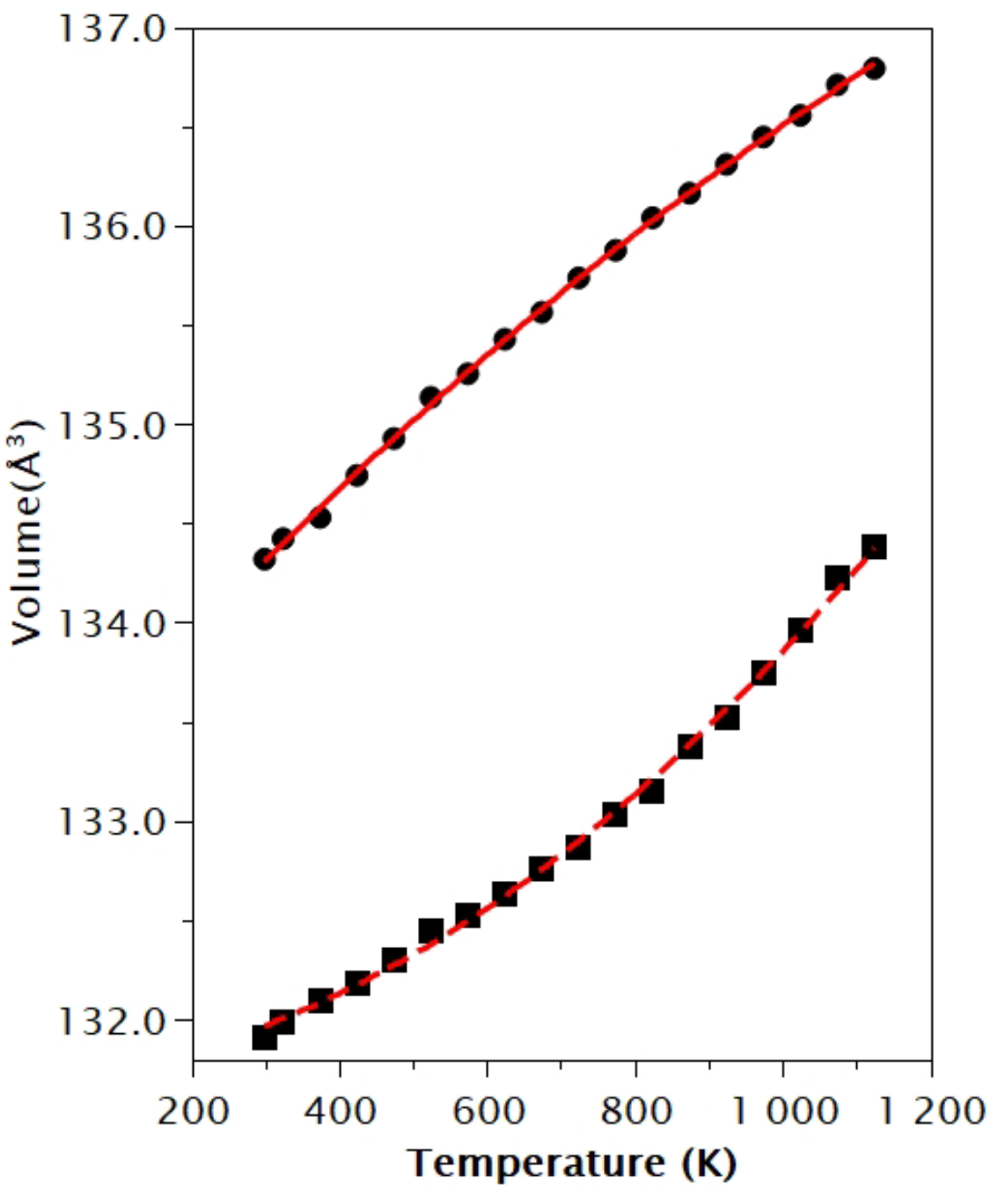

**Figure 6.** High temperature behavior of ferberite (circles) and synthetic $FeWO_4$ (squares). Error bars are smaller than symbols. The lines show the fits of the Berman model to the data.

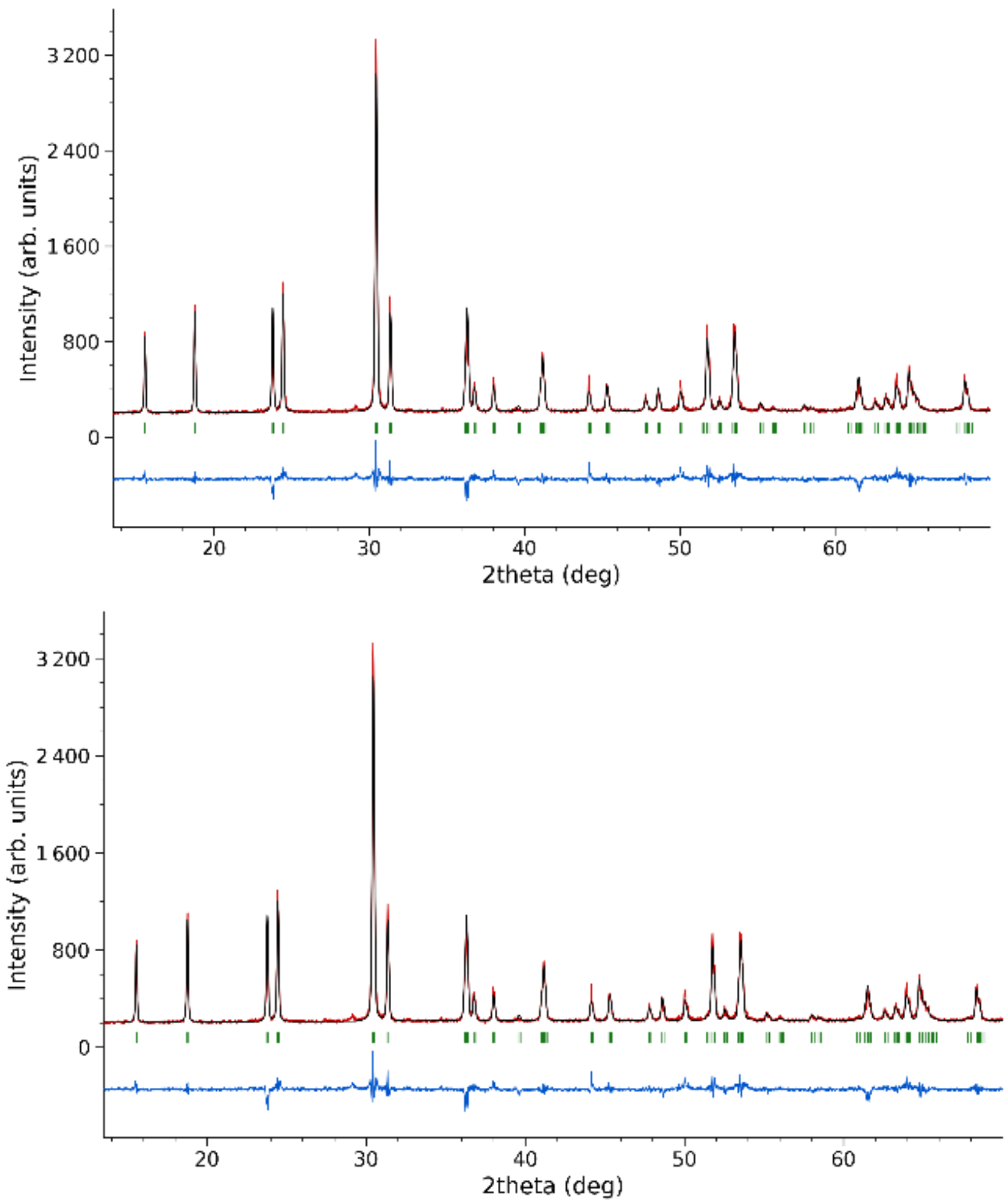

**Figure 7**: Rietveld refinement of the X-ray powder diffraction patterns of ferberite measured on an X'Pert diffractometer using Cu K$_\alpha$ radiation at 300 K, before (Top) and after (Bottom) heating to 1123 K. The experimental data and calculated profiles are shown as red and black lines, respectively, while the blue line represents the difference curve. Vertical green ticks indicate the positions of the Bragg reflections. See details on Table 3.

## 4.- Conclusions

The thermal expansion of $FeWO_4$ has been systematically investigated over an exceptionally broad temperature interval (2–1123 K) by combining single-crystal X-ray diffraction, powder X-ray diffraction, and neutron powder diffraction. This multi-technique approach enables a robust and internally consistent determination of the temperature dependence of the unit-cell volume across both low- and high-temperature regimes.

In the 2–523 K range, the experimental data are accurately described using the physically based model of Kroll *et al.*, which properly accounts for the low-temperature saturation of the thermal expansion coefficient and yields thermodynamically meaningful parameters. At higher temperatures (298–1123 K), the formulation proposed by Berman successfully reproduces the observed non-linear evolution of volume with temperature. The consistency between both approaches confirms the thermodynamic stability of $FeWO_4$ throughout the investigated range and indicates that no structural phase transition occurs up to 1123 K under ambient pressure.

A systematic difference is observed between the natural $FeWO_4$ (ferberite) sample and the $FeWO_4$:$Fe_2WO_6$ (7:1) composite (synthetic $FeWO_4$). The composite exhibits a reference unit-cell volume about 2% smaller and a thermal expansion coefficient approximately 40% lower than that of phase-pure $FeWO_4$. Instrumental artefacts and simple compositional effects can reasonably be excluded as the origin of this discrepancy. Instead, the results point to intrinsic multiphase and microstructural effects. In particular, the presence of the $Fe_2WO_6$ secondary phase likely introduces lattice strain and mechanical constraints within the composite, effectively limiting the apparent expansion of the $FeWO_4$ phase.

A secondary contribution could arise from partial oxidation of $Fe^{2+}$ to $Fe^{3+}$ within the $FeWO_4$ lattice, for instance due to slight over-oxidation during synthesis or thermal treatment. Such a substitution would reduce the effective ionic radius of iron, leading to shorter Fe–O bonds and a locally stiffer lattice, which may decrease the thermal expansion coefficient. The Fe–O distances obtained from the single-crystal refinement of natural $FeWO_4$ (2.06–2.13–2.18 Å) are consistent with $Fe^{2+}$ in octahedral coordination and agree, within experimental uncertainty, with previously reported values. In contrast, the Fe–O distances derived from the Rietveld refinement of the synthesized sample measured at room temperature on the D20 neutron diffractometer are slightly shorter (2.03–2.11–2.17 Å), which could be compatible with the presence of a small $Fe^{3+}$ fraction. The occurrence of a minor $Fe_2WO_6$ phase in the synthesized sample, where iron is stabilized in the $Fe^{3+}$ state, further suggests that local oxidation processes may occur during synthesis, making a partial $Fe^{2+} \rightarrow Fe^{3+}$ substitution within the $FeWO_4$ lattice plausible. Nevertheless, this scenario remains speculative and would require a direct determination of the iron oxidation state in the $FeWO_4$ phase. In practice, the coexistence of $Fe_2WO_6$ complicates such analyses and prevents an unambiguous determination using techniques such as x-ray absorption spectroscopy, Mössbauer spectroscopy, or quantitative x-ray photoelectron spectroscopy.

Overall, this work provides a comprehensive thermoelastic description of $FeWO_4$ and demonstrates that derived thermal expansion parameters can be significantly influenced by phase coexistence and microstructural conditions. The study highlights the importance of combining complementary diffraction techniques and carefully characterizing sample composition when determining reliable thermoelastic properties in complex transition-metal oxides.

## Acknowledgments

The authors would like to thank Institut Laue Langevin (ILL) for access to beamtime in proposals 5-25-294 and Easy1507.

## Conflict of Interest

We declare no conflict of interest in this work.

## Data Availability

**Data availability**

Neutron diffraction data are deposited in the Institut Laue-Langevin (ILL) and are openly available in https://doi.ill.fr/10.5291/ILL-DATA.5-25-294 and https://doi.ill.fr/10.5291/ILL-DATA.EASY-1507. A complete crystallographic information of the structures can be obtained from the Cambridge Crystallographic Data Centre (CCDC) under deposition numbers 2551363 – 2551375. The rest of the data is available in the Supporting Information.

## Supporting Information

Schematic representation of the crystal structure. Detail information of crystal structure and structural refinements at different conditions. Bond distances versus temperature and unit-cell parameters measured under different conditions.

## Funding

This work has been supported by grant ProID2024010034 funded by the Agencia Canaria de Investigación, Innovación y Sociedad de la Información (ACIISI) and by the Fondo Europeo de Desarrollo Regional en el marco del programa FEDER Canarias 2021-2027. It was also supported by grant PID2024-158791NB-I00 from NICIU/AEI/10.13039/501100011033. D.E. thanks the support from Ministerio de Ciencia, Innovación y Universidades (MCIU) from Spain, through Agencia Estatal de investigación (DOI: MCIN/AEI/10.13039/501100011033) under grant PID2022-138076NB-C41.

## Bibliography

[1] Muñoz, A., Radescu, S., Mujica, A., & Errandonea, D. (2026) J. Phys. Chem. 130, 9, 3201–3225.

[2] Cid-Dresdner, H. (1968) Zeitschrift für Kristallographie 127, 61-72.

[3] Diaz-Anichtchenko, D., Aviles-Coronado, J. E., López-Moreno, S., Turnbull, R., Manjón, F. J., Popescu, C., & Errandonea D. (2024) Inorg. Chem. 63, 6898-6908.

[4] Heyer, O., Hollmann, N., Klassen, I., Jodlauk, S., Bohatý, L., Becker, P., Mydosh, J. A., Lorenz, T., & Khomskii, D. (2006= J. Phys.: Condens. Matter 18, L471−L475.

[5] García-Matres, E., Stüßer, N., Hofmann, M., & Reehuis, M. (2003) Eur. Phys. J. B 32, 35−42.

[6] Maignan, A., Schmidt, M., Prots, Y., Lebedev, O. I., Daou, R., Chang, C. F., Kuo, C. Y., Hu, Z., Chen, C. T., Weng, C. C., Altendorf, S. G., Tjeng, L. H., & Grin, Y. (2022) Chem. Mat. 34, 789-797.

[7] Kroll H., Kirfel A., Heinemann R., & Barbier B. (2012) Eur. J. Miner. 24, 935-956.

[8] Salje E. K. H., Wruck B., & Thomas H. (1991) Zeitschrift für Physik B 82, 399-404.

[9] Berman R. G. (1988) J. Petrol. 29, 445-522.

[10] Fabelo, O., Gonzalez-Platas, J., Savvin, S., Botella, P., & Errandonea, D. (2024) J. Appl. Phys. 136, 175901.

[11] Gonzalez-Platas, J., Alvaro, M., Nestola, F., Angel, R. 2016, J. Appl. Cryst, 49, 1377-1382.

[12] Bruker. SAINT. Version 8.34a. Bruker AXS Inc. Madison, Wisconsin, EEUU (2014).

[13] Le Bail, A., Duroy, H., & Fourquet, J. L. (1998) Mater. Res. Bull. 23, 447-452.

[14] Rodríguez-Carvajal, J. (1993) Physica B 192, 55-69.

[15] T. C. Hansen, P. F. Henry, H. E. Fischer, J. Torregrossa, and P. Convert, Meas. Sci. Technol., 2008, 19, 034001

[16] Finger, L. W., Cox, D. E., & Jephcoat, A. P. (1994) J. Appl. Cryst. 27, 892–900.

[17] Available in the online ILL repository, https://code.ill.fr/fabelo/panda

[18] Qin, W., Nagase, T., Umakoshi, Y., & Szpunar, J. A. (2008) Phil. Mag. Letters 88, 169–179.

[19] Akhlaghi, M., Steiner, T., Meka, S. R. & Mittemeijer, E. J. (2016) J. Appl. Cryst. 49, 69-77.

[20] Hsieh, C. L., & Tuan, W.H. & Mater. Sci. Eng. A 425, 349–360.

[21] Takeda, M., Onishi, T., Nakakubo, S., & Fujimoto, S. (2009) Mat. Trans. 50, 2242-2246.

[22] Zou, Y., Wang, P., Li, Y., Chen, H., Zhou, C., & Irifune, T, (2025) iScience 28, 111905.

[23] Yang, G., Liu, X., Sun, X., Liang, E., & Zhang, W. (2018) Ceram. Int. 44, 22032-22035.

[24] Shannon, R. D. (1976) Acta Crystallogr. A 32, 751–767.

[25] Errandonea, D., & Manjon, F. J. (2008) Progress in Materials Science 53, 711-773.

[26] Dubrovinskaia, N. A., Dubrovinsky, L. S., Saxena, S. K., & Sundman, B. (1997) Calphad 21, 497-508.

[27] Drebushchak, V. A. (2020) J. Thermal Anal. Calor. 142, 1097–1113.

# SUPPORTING INFORMATION

# Thermal expansion of $FeWO_4$ (Ferberite) and $FeWO_4$:$Fe_2WO_6$ (7:1): a comparative X-ray and neutron diffraction study

O. Fabelo[1], L. Cañadillas-Delgado[1], D. Vie[2], E. Matesanz[3], J. Gonzalez-Platas[4], and D. Errandonea[5,*]

[1] Institut Laue-Langevin, 71 avenue des Martyrs, CS 20156, 38042 Grenoble cedex 9, France

[2] Institut de Ciència dels Materials de la Universitat de València, Apartado de Correos 2085, E-46071 València, Spain

[3] Unidad de Difracción de Rayos X, Centro de asistencia a la Investigación de Técnicas Químicas, Universidad Complutense de Madrid, Madrid, Spain

[4] Departamento de Física, Instituto Universitario de estudios Avanzados en Física Atómica, Molecular y Fotónica (IUDEA), MALTA Consolider Team, Universidad de La Laguna, La Laguna, Tenerife, Spain

[5] Departamento de Física Aplicada-ICMUV, MALTA Consolider Team, Universitat de Valencia, Dr Moliner 50, Burjassot, Valencia, Spain

*Corresponding author: daniel.errandonea@uv.es

## Ferberite crystal structure

**Figure S1:** Schematic representation of Ferberite structure. Tungsten (iron) atoms and shown in grey (brown) and oxygen atoms in red. The WO6 and FeO6 octahedra are represented.

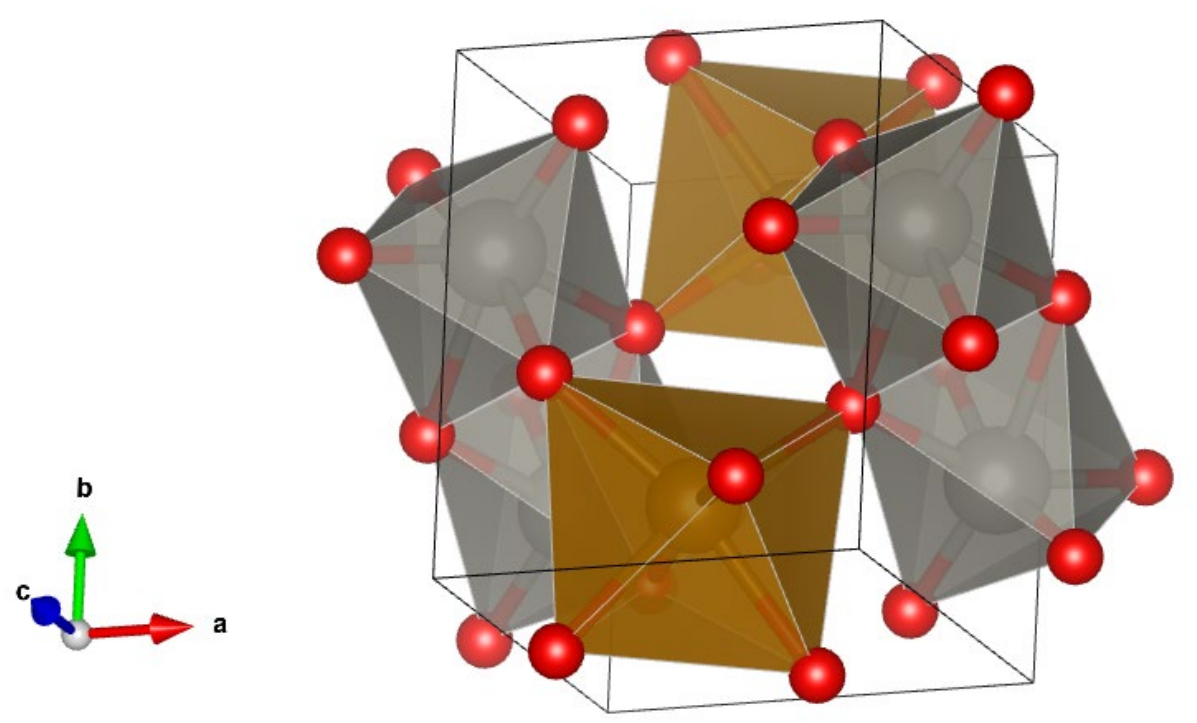

## Ferberite - Single Crystal Diffraction

**Table S1.** Crystallographic data and structure refinement for Ferberite

| Empirical formula | $FeWO_4$ | $FeWO_4$ |
|---|---|---|
| Formula weight | 303.685 | 303.685 |
| *T*/K | **40.0(5)** | **60.0(5)** |
| λ/Å | 0.56087 | 0.56087 |
| Crystal system | Monoclinic | Monoclinic |
| Space group | *P2/c* | *P2/c* |
| *a* /Å | 4.7358(3) | 4.7352(3) |
| *b* /Å | 5.7092(3) | 5.7100(3) |
| *c* /Å | 4.9562(3) | 4.9573(3) |
| α /° | 90 | 90 |
| β /° | 90.156(4) | 90.154(3) |
| γ /° | 90 | 90 |
| Volume /Å$^3$ | 134.003(14) | 134.035(14) |
| *Z* | 2 | 2 |
| $\rho_{calcd}$/g cm$^{-3}$ | 7.526 | 7.525 |
| μ/mm$^{-1}$ | 25.803 | 25.797 |
| *F*(000) | 264.3 | 264.3 |
| Crystal size /mm$^3$ | 0.10 x 0.07 x 0.05 | 0.10 x 0.07 x 0.05 |
| 2θ range for data collection/° | 5.64 to 68.24 | 5.64 to 68.28 |
| Index ranges | $-6 \le h \le 9$<br>$-11 \le k \le 11$<br>$-9 \le l \le 9$ | $-6 \le h \le 9$<br>$-11 \le k \le 11$<br>$-9 \le l \le 9$ |
| Reflect. collected | 5012 | 5235 |
| Independent reflect. [*R*(int)] | 1119 [0.0805] | 1116 [0.0773] |
| Data/restraints/ parameters | 1119/0/30 | 1116/0/30 |
| Goodness-of-fit on $F^2$ | 1.078 | 0.956 |
| Final *R* indices [$I > 2\sigma(I)$] | $R_1 = 0.0819$, $wR_2 = 0.2086$ | $R_1 = 0.0709$, $wR_2 = 0.1818$ |
| *R* indices (all data) | $R_1 = 0.0840$, $wR_2 = 0.2144$ | $R_1 = 0.0725$, $wR_2 = 0.1853$ |
| Largest diff. peak and hole/e Å$^{-3}$ | 1.16 /-1.30 | 1.21 /-1.41 |

**Table S1 (continuation).** Crystallographic data and structure refinement for Ferberite

| Empirical formula | $FeWO_4$ | $FeWO_4$ |
|---|---|---|
| Formula weight | 303.685 | 303.685 |
| *T*/K | **80.0(5)** | **100.0(5)** |
| λ/Å | 0.56087 | 0.56087 |
| Crystal system | Monoclinic | Monoclinic |
| Space group | *P*2/*c* | *P*2/*c* |
| *a* /Å | 4.7338(5) | 4.7361(6) |
| *b* /Å | 5.7111(6) | 5.7078(7) |
| *c* /Å | 4.9577(6) | 4.9590(6) |
| α /° | 90 | 90 |
| β /° | 90.147(4) | 90.094(6) |
| γ /° | 90 | 90 |
| Volume /$Å^3$ | 134.03(14) | 134.06(3) |
| *Z* | 2 | 2 |
| $\rho_{calcd}$/g $cm^{-3}$ | 7.525 | 7.523 |
| μ/$mm^{-1}$ | 25.797 | 25.793 |
| *F*(000) | 264.3 | 264.3 |
| Crystal size /$mm^3$ | 0.10 x 0.07 x 0.05 | 0.10 x 0.07 x 0.05 |
| 2θ range for data collection/° | 5.62 to 68.26 | 5.64 to 68.26 |
| Index ranges | $-6 \leq h \leq 9$<br>$-11 \leq k \leq 11$<br>$-9 \leq l \leq 9$ | $-6 \leq h \leq 9$<br>$-11 \leq k \leq 11$<br>$-9 \leq l \leq 9$ |
| Reflect. collected | 5412 | 5545 |
| Independent reflect. [*R*(int)] | 1119 [0.0766] | 1118 [0.0742] |
| Data/restraints/ parameters | 1119/0/30 | 1118/0/30 |
| Goodness-of-fit on $F^2$ | 0.963 | 0.907 |
| Final *R* indices [$I > 2\sigma(I)$] | $R_1 = 0.0567$, $wR_2 = 0.1522$ | $R_1 = 0.0540$, $wR_2 = 0.1423$ |
| *R* indices (all data) | $R_1 = 0.0585$, $wR_2 = 0.1550$ | $R_1 = 0.0567$, $wR_2 = 0.1536$ |
| Largest diff. peak and hole/e $Å^{-3}$ | 0.86 /-0.75 | 0.73 /-0.90 |

**Table S1 (continuation).** Crystallographic data and structure refinement for Ferberite

| Empirical formula | $FeWO_4$ | $FeWO_4$ |
|---|---|---|
| Formula weight | 303.685 | 303.685 |
| $T$/K | **130.0(5)** | **160.0(5)** |
| λ/Å | 0.56087 | 0.56087 |
| Crystal system | Monoclinic | Monoclinic |
| Space group | *P*2/*c* | *P*2/*c* |
| *a* /Å | 4.7365(5) | 4.7375(5) |
| *b* /Å | 5.7073(7) | 5.7074(6) |
| *c* /Å | 4.9614(6) | 4.9622(5) |
| α /° | 90 | 90 |
| β /° | 90.073(4) | 90.057(5) |
| γ /° | 90 | 90 |
| Volume /$Å^3$ | 134.12(3) | 134.12(3) |
| *Z* | 2 | 2 |
| $\rho_{calcd}$/g $cm^{-3}$ | 7.526 | 7.525 |
| μ/$mm^{-1}$ | 25.780 | 25.780 |
| *F*(000) | 264.3 | 264.3 |
| Crystal size /$mm^3$ | 0.10 x 0.07 x 0.05 | 0.10 x 0.07 x 0.05 |
| 2θ range for data collection/° | 5.64 to 68.24 | 5.64 to 68.24 |
| Index ranges | $-6 \leq h \leq 9$<br>$-11 \leq k \leq 11$<br>$-9 \leq l \leq 9$ | $-6 \leq h \leq 9$<br>$-11 \leq k \leq 11$<br>$-9 \leq l \leq 9$ |
| Reflect. collected | 5417 | 5318 |
| Independent reflect. [*R*(int)] | 1120 [0.0716] | 1122 [0.0730] |
| Data/restraints/ parameters | 1120/0/30 | 1122/0/30 |
| Goodness-of-fit on $F^2$ | 0.959 | 0.931 |
| Final *R* indices [$I > 2\sigma(I)$] | $R_1 = 0.0522$, $wR_2 = 0.1364$ | $R_1 = 0.0550$, $wR_2 = 0.1480$ |
| *R* indices (all data) | $R_1 = 0.0532$, $wR_2 = 0.1381$ | $R_1 = 0.0559$, $wR_2 = 0.1497$ |
| Largest diff. peak and hole/e $Å^{-3}$ | 1.0 /-0.82 | 0.97 /-0.82 |

**Table S1 (continuation).** Crystallographic data and structure refinement for Ferberite

| Empirical formula | $FeWO_4$ | $FeWO_4$ |
|---|---|---|
| Formula weight | 303.685 | 303.685 |
| $T$/K | **200.0(5)** | **250.0(5)** |
| $\lambda$/Å | 0.56087 | 0.56087 |
| Crystal system | Monoclinic | Monoclinic |
| Space group | *P2/c* | *P2/c* |
| $a$ /Å | 4.7397(6) | 4.7394(3) |
| $b$ /Å | 5.7082(7) | 5.7125(4) |
| $c$ /Å | 4.9647(6) | 4.9686(3) |
| $\alpha$ /° | 90 | 90 |
| $\beta$ /° | 90.028(7) | 90.004(4) |
| $\gamma$ /° | 90 | 90 |
| Volume /Å$^3$ | 134.32(3) | 134.519(15) |
| $Z$ | 2 | 2 |
| $\rho_{calcd}$/g cm$^{-3}$ | 7.509 | 7.498 |
| $\mu$/mm$^{-1}$ | 25742 | 25.704 |
| $F$(000) | 264.3 | 264.3 |
| Crystal size /mm$^3$ | 0.10 x 0.07 x 0.05 | 0.10 x 0.07 x 0.05 |
| 2θ range for data collection/° | 5.64 to 68.20 | 5.62 to 68.16 |
| Index ranges | $-6 \leq h \leq 9$<br>$-11 \leq k \leq 11$<br>$-9 \leq l \leq 9$ | $-6 \leq h \leq 9$<br>$-11 \leq k \leq 11$<br>$-9 \leq l \leq 9$ |
| Reflect. collected | 5447 | 5552 |
| Independent reflect. [$R$(int)] | 1126 [0.0788] | 1124 [0.0819] |
| Data/restraints/ parameters | 1126/0/30 | 1124/0/30 |
| Goodness-of-fit on $F^2$ | 0.934 | 0.948 |
| Final $R$ indices [$I > 2\sigma(I)$] | $R_1$ = 0.0614, w$R_2$ = 0.1562 | $R_1$ = 0.0514, w$R_2$ = 0.1373 |
| $R$ indices (all data) | $R_1$ = 0.0626, w$R_2$ = 0.1586 | $R_1$ = 0.0534, w$R_2$ = 0.1403 |
| Largest diff. peak and hole/e Å$^{-3}$ | 0.72 /-1.12 | 0.86 /-0.83 |

**Table S1 (continuation).** Crystallographic data and structure refinement for Ferberite

| Empirical formula | $FeWO_4$ | $FeWO_4$ |
|---|---|---|
| Formula weight | 303.685 | 303.685 |
| $T$/K | **300.0(5)** | **350.0(5)** |
| $\lambda$/Å | 0.56087 | 0.56087 |
| Crystal system | Monoclinic | Monoclinic |
| Space group | $P2/c$ | $P2/c$ |
| $a$ /Å | 4.7418(1) | 4.7425(2) |
| $b$ /Å | 5.7156(2) | 5.7174(2) |
| $c$ /Å | 4.9712(2) | 4.9733(2) |
| $\alpha$ /° | 90 | 90 |
| $\beta$ /° | 89.9660(17) | 89.924 (2) |
| $\gamma$ /° | 90 | 90 |
| Volume /Å$^3$ | 134.731(8) | 134.850(9) |
| $Z$ | 2 | 2 |
| $\rho_{calcd}$/g cm$^{-3}$ | 7.486 | 7.479 |
| $\mu$/mm$^{-1}$ | 25.663 | 25.641 |
| $F$(000) | 264.3 | 264.3 |
| Crystal size /mm$^3$ | 0.10 x 0.07 x 0.05 | 0.10 x 0.07 x 0.05 |
| 2θ range for data collection/° | 5.62 to 68.22 | 5.62 to 68.22 |
| Index ranges | $-6 \leq h \leq 9$<br>$-11 \leq k \leq 11$<br>$-9 \leq l \leq 9$ | $-6 \leq h \leq 9$<br>$-11 \leq k \leq 11$<br>$-9 \leq l \leq 9$ |
| Reflect. collected | 5621 | 5664 |
| Independent reflect. [$R$(int)] | 1124 [0.0634] | 1126 [0.0701] |
| Data/restraints/ parameters | 1124/0/30 | 1126/0/30 |
| Goodness-of-fit on $F^2$ | 0.971 | 0.972 |
| Final $R$ indices [$I > 2\sigma(I)$] | $R_1$ = 0.0447, $wR_2$ = 0.1202 | $R_1$ = 0.0459, $wR_2$ = 0.1212 |
| $R$ indices (all data) | $R_1$ = 0.0457, $wR_2$ = 0.1215 | $R_1$ = 0.0472, $wR_2$ = 0.1225 |
| Largest diff. peak and hole/e Å$^{-3}$ | 0.72 /-0.60 | 0.75 /-0.55 |

**Table S1 (continuation).** Crystallographic data and structure refinement for Ferberite

| Empirical formula | $FeWO_4$ | $FeWO_4$ |
|---|---|---|
| Formula weight | 303.685 | 303.685 |
| *T*/K | **400.0(5)** | **450.0(5)** |
| λ/Å | 0.56087 | 0.56087 |
| Crystal system | Monoclinic | Monoclinic |
| Space group | *P*2/*c* | *P*2/*c* |
| *a* /Å | 4.7432(2) | 4.7461(3) |
| *b* /Å | 5.7195(3) | 5.7245(4) |
| *c* /Å | 4.9754(2) | 4.9784(3) |
| α /° | 90 | 90 |
| β /° | 89.887 (3) | 89.838(4) |
| γ /° | 90 | 90 |
| Volume /Å$^3$ | 134.976(11) | 135.258(15) |
| *Z* | 2 | 2 |
| $\rho_{calcd}$/g cm$^{-3}$ | 7.472 | 7.457 |
| μ/mm$^{-1}$ | 25.617 | 25.563 |
| *F*(000) | 264.3 | 264.3 |
| Crystal size /mm$^3$ | 0.10 x 0.07 x 0.05 | 0.10 x 0.07 x 0.05 |
| 2θ range for data collection/° | 5.62 to 68.14 | 5.62 to 68.24 |
| Index ranges | $-6 \leq h \leq 9$<br>$-11 \leq k \leq 11$<br>$-9 \leq l \leq 9$ | $-6 \leq h \leq 9$<br>$-11 \leq k \leq 11$<br>$-9 \leq l \leq 9$ |
| Reflect. collected | 5619 | 5678 |
| Independent reflect. [*R*(int)] | 1122 [0.0739] | 1127 [0.0751] |
| Data/restraints/ parameters | 1122/0/30 | 1127/0/30 |
| Goodness-of-fit on $F^2$ | 0.974 | 0.967 |
| Final *R* indices [$I > 2\sigma(I)$] | $R_1$ = 0.0463, $wR_2$ = 0.1210 | $R_1$ = 0.0445, $wR_2$ = 0.1201 |
| *R* indices (all data) | $R_1$ = 0.0479, $wR_2$ = 0.1230 | $R_1$ = 0.0466, $wR_2$ = 0.1232 |
| Largest diff. peak and hole/e Å$^{-3}$ | 0.62 /-0.59 | 0.56 /-0.52 |

**Table S1 (continuation).** Crystallographic data and structure refinement for Ferberite

| Empirical formula | $FeWO_4$ |
|---|---|
| Formula weight | 303.685 |
| $T$/K | **500.0(5)** |
| $\lambda$/Å | 0.56087 |
| Crystal system | Monoclinic |
| Space group | $P2/c$ |
| $a$ /Å | 4.7474(4) |
| $b$ /Å | 5.7266(5) |
| $c$ /Å | 4.9798 (4) |
| $\alpha$ /° | 90 |
| $\beta$ /° | 89.790(5) |
| $\gamma$ /° | 90 |
| Volume /Å$^3$ | 135.38(2) |
| $Z$ | 2 |
| $\rho_{calcd}$/g cm$^{-3}$ | 7.450 |
| $\mu$/mm$^{-1}$ | 25.540 |
| $F$(000) | 264.3 |
| Crystal size /mm$^3$ | 0.10 x 0.07 x 0.05 |
| 2θ range for data collection/° | 5.62 to 68.00 |
| Index ranges | $-6 \leq h \leq 9$<br>$-11 \leq k \leq 11$<br>$-9 \leq l \leq 9$ |
| Reflect. collected | 5049 |
| Independent reflect. [$R$(int)] | 1107 [0.0745] |
| Data/restraints/ parameters | 1107/0/30 |
| Goodness-of-fit on $F^2$ | 0.974 |
| Final $R$ indices [$I > 2\sigma(I)$] | $R_1 = 0.0549$, $wR_2 = 0.1485$ |
| $R$ indices (all data) | $R_1 = 0.0591$, $wR_2 = 0.1565$ |
| Largest diff. peak and hole/e Å$^{-3}$ | 0.82 /-0.83 |

**Ferberite Sample – SC-Xray**

**Table S2:** Distances between W-O and Fe-O vs function of temperature.

| T(K) | W-O1 | W-O2 | W-O1a | Fe-O1 | Fe-O2 | Fe-O2a |
|---|---|---|---|---|---|---|
| 40.0(5) | 1.908(7) | 1.796(7) | 2.126(7) | 2.068(7) | 2.134(7) | 2.169(7) |
| 60.0(5) | 1.909(6) | 1.795(6) | 2.125(6) | 2.070(6) | 2.135(6) | 2.173(6) |
| 80.0(5) | 1.914(5) | 1.789(6) | 2.128(5) | 2.061(5) | 2.135(5) | 2.182(6) |
| 100.0(5) | 1.912(5) | 1.795(5) | 2.132(5) | 2.062(5) | 2.129(5) | 2.177(5) |
| 130.0(5) | 1.910(4) | 1.795(5) | 2.130(5) | 2.065(5) | 2.134(4) | 2.172(5) |
| 160.0(5) | 1.913(4) | 1.793(5) | 2.133(5) | 2.061(5) | 2.133(5) | 2.179(5) |
| 200.0(5) | 1.912(5) | 1.797(5) | 2.128(5) | 2.069(5) | 2.131(5) | 2.178(6) |
| 250.0(5) | 1.911(4) | 1.789(5) | 2.129(5) | 2.069(5) | 2.138(5) | 2.186(5) |
| 300.0(5) | 1.916(4) | 1.792(5) | 2.133(5) | 2.065(4) | 2.134(4) | 2.189(5) |
| 350.0(5) | 1.917(4) | 1.791(5) | 2.133(4) | 2.063(5) | 2.138(4) | 2.190(5) |
| 400.0(5) | 1.914(4) | 1.793(5) | 2.131(5) | 2.069(5) | 2.141(4) | 2.188(5) |
| 450.0(5) | 1.920(5) | 1.788(5) | 2.137(5) | 2.064(5) | 2.141(5) | 2.197(5) |
| 500.0(5) | 1.911(5) | 1.792(5) | 2.134(6) | 2.075(6) | 2.141(6) | 2.198(6) |

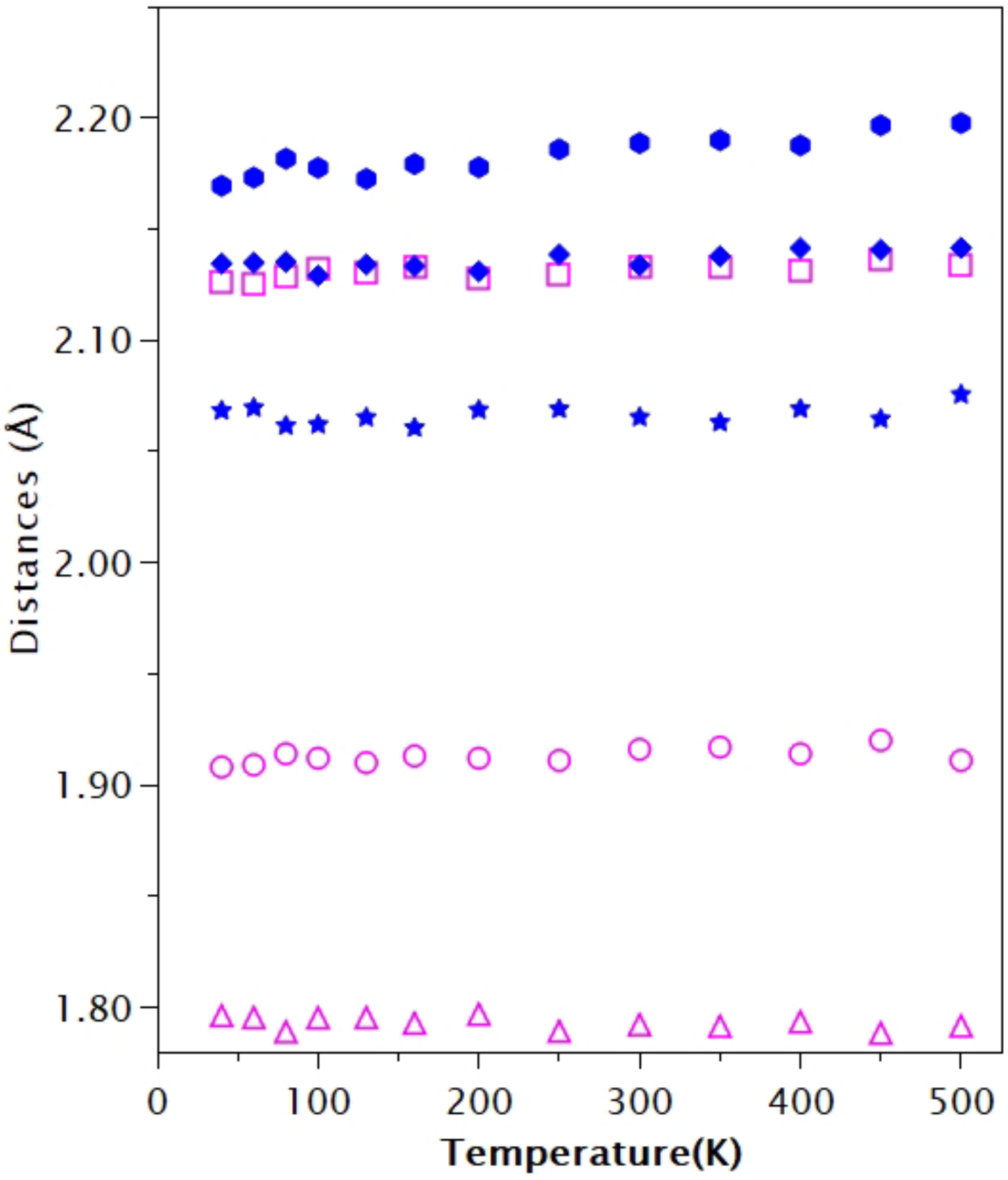

**Figure S2:** Distances evolution with temperature for Fe-O (blue fill symbols) and W-O (magenta empty symbols). Error bars are smaller than the size symbols.

**Ferberite Sample – SC-Xray**

**Table S3:** Unit-cell parameters evolution with temperature.

| T(K) | *a(Å)* | *b(Å)* | *c(Å)* | β(º) |
|---|---|---|---|---|
| 40.0(5) | 4.7358(3) | 5.7092(3) | 4.9562(3) | 90.156(4) |
| 60.0(5) | 4.7352(3) | 5.7100(3) | 4.9573(3) | 90.154(4) |
| 80.0(5) | 4.7338(5) | 5.7111(6) | 4.9577(6) | 90.147(4) |
| 100.0(5) | 4.7361(6) | 5.7078(7) | 4.9590(6) | 90.094(6) |
| 130.0(5) | 4.7365(5) | 5.7073(7) | 4.9614(5) | 90.073(4) |
| 160.0(5) | 4.7375(5) | 5.7074(6) | 4.9622(5) | 90.057(5) |
| 200.0(5) | 4.7397(6) | 5.7082(7) | 4.9647(6) | 90.028(7) |
| 250.0(5) | 4.7394(3) | 5.7125(4) | 4.9686(3) | 90.004(4) |
| 300.0(5) | 4.7418(2) | 5.7156(2) | 4.9712(2) | 89.966(2) |
| 350.0(5) | 4.7425(2) | 5.7174(2) | 4.9733(2) | 89.924(2) |
| 400.0(5) | 4.7432(2) | 5.7195(3) | 4.9754(2) | 89.887(3) |
| 450.0(5) | 4.7461(3) | 5.7245(4) | 4.9784(3) | 89.838(4) |
| 500.0(5) | 4.7474(4) | 5.7266(5) | 4.9798(4) | 89.790(5) |

**Ferberite Sample – XRD**

**Table S4:** Unit-cell parameters evolution with temperature from 15-523 K.

| T(K) | a*(Å)* | b*(Å)* | c*(Å)* | β(º) |
|---|---|---|---|---|
| 15(1) | 4.73425(13) | 5.70661(11) | 4.94901(14) | 90.254(3) |
| 35(1) | 4.73413(13) | 5.70706(11) | 4.94839(14) | 90.255(3) |
| 45(1) | 4.73377(14) | 5.70698(11) | 4.94909(14) | 90.235(3) |
| 55(1) | 4.73432(14) | 5.70766(12) | 4.94949(14) | 90.242(3) |
| 65(1) | 4.73395(13) | 5.70765(11) | 4.94977(14) | 90.259(3) |
| 75(1) | 4.73371(14) | 5.70807(11) | 4.95063(14) | 90.252(3) |
| 85(1) | 4.73412(13) | 5.70801(11) | 4.95154(14) | 90.234(3) |
| 100(1) | 4.73421(13) | 5.70828(12) | 4.95169(14) | 90.249(3) |
| 115(1) | 4.73451(13) | 5.70859(12) | 4.95191(14) | 90.254(3) |
| 130(1) | 4.73508(14) | 5.70849(14) | 4.95360(15) | 90.198(3) |
| 145(1) | 4.73492(14) | 5.70824(11) | 4.95333(15) | 90.177(3) |
| 160(1) | 4.73558(13) | 5.70832(12) | 4.95488(15) | 90.176(3) |
| 180(1) | 4.73676(13) | 5.70877(13) | 4.95705(16) | 90.114(4) |
| 200(1) | 4.73688(13) | 5.70913(11) | 4.95680(15) | 90.140(3) |
| 220(1) | 4.73826(13) | 5.71014(12) | 4.96059(15) | 90.178(3) |
| 240(1) | 4.73864(13) | 5.71050(12) | 4.96061(15) | 90.150(3) |
| 280(1) | 4.73975(13) | 5.71122(11) | 4.96025(15) | 90.128(3) |
| 290(1) | 4.73975(13) | 5.71176(12) | 4.96384(15) | 90.160(3) |
| 298(1) | 4.74053(14) | 5.71213(14) | 4.96472(15) | 90.154(5) |
| 300(1) | 4.74053(12) | 5.71213(12) | 4.96472(15) | 90.154(3) |
| 323(1) | 4.74262(25) | 5.71266(25) | 4.96520(29) | 90.145(6) |
| 373(1) | 4.74313(27) | 5.71382(26) | 4.96457(27) | 90.221(6) |
| 423(1) | 4.74569(25) | 5.71690(25) | 4.96664(24) | 90.317(5) |
| 473(1) | 4.74769(24) | 5.71887(25) | 4.96884(23) | 90.347(4) |
| 523(1) | 4.74912(23) | 5.72079(25) | 4.97096(26) | 90.339(5) |

**LeBail Refinements**

| T(K) | $R_p$ | $R_{wp}$ | $R_{exp}$ | $\chi^2$ | $R_{Bragg}$ |
|---|---|---|---|---|---|
| 15-523 | 14.8-15.7 | 19.5-21.1 | 11.88-12.34 | 2.70-2.99 | 1.93-2.38 |

**Table S5:** Unit-cell parameters evolution with temperature from 298-1123 K.

| T(K) | a(Å) | b(Å) | c(Å) | β(º) |
|---|---|---|---|---|
| 298(1) | 4.7377(10) | 5.7104(11) | 4.9650(12) | 90.087(4) |
| 323(1) | 4.7405(12) | 5.7115(12) | 4.9649(13) | 90.169(3) |
| 373(1) | 4.7415(12) | 5.7129(12) | 4.9666(14) | 90.184(3) |
| 423(1) | 4.7441(12) | 5.7156(12) | 4.9694(14) | 90.202(3) |
| 473(1) | 4.7465(12) | 5.7175(12) | 4.9720(14) | 90.238(2) |
| 523(1) | 4.7488(12) | 5.7200(12) | 4.9751(14) | 90.265(2) |
| 573(1) | 4.7505(12) | 5.7219(12) | 4.9760(14) | 90.294(18) |
| 623(1) | 4.7523(12) | 5.7243(12) | 4.9785(14) | 90.321(18) |
| 673(1) | 4.7539(12) | 5.7263(12) | 4.9801(14) | 90.355(17) |
| 723(1) | 4.7561(12) | 5.7283(12) | 4.9824(14) | 90.384(17) |
| 773(1) | 4.7577(13) | 5.7305(12) | 4.9839(15) | 90.409(16) |
| 823(1) | 4.7597(13) | 5.7329(12) | 4.9858(15) | 90.437(16) |
| 873(1) | 4.7614(13) | 5.7348(12) | 4.9870(15) | 90.463(16) |
| 923(1) | 4.7629(13) | 5.7373(12) | 4.9885(14) | 90.495(15) |
| 973(1) | 4.7645(13) | 5.7388(12) | 4.9906(15) | 90.519(15) |
| 1023(1) | 4.7656(12) | 5.7407(12) | 4.9917(15) | 90.538(15) |
| 1073(1) | 4.7675(12) | 5.7432(11) | 4.9933(13) | 90.554(15) |
| 1123(1) | 4.7683(12) | 5.7444(12) | 4.9944(14) | 90.582(15) |

**LeBail Refinements**

| T(K) | $R_p$ | $R_{wp}$ | $R_{exp}$ | $\chi^2$ | $R_{Bragg}$ |
|---|---|---|---|---|---|
| 298-1123 | 13.6-16.1 | 17.8-21.0 | 10.56-10.68 | 2.79-3.92 | 1.74-2.22 |

**$FeWO_4$ Synthesized Sample – XRD**

**Table S6:** Unit-cell parameters evolution with temperature from 15-523 K.

| T(K) | a(Å) | b(Å) | c(Å) | β(º) |
|---|---|---|---|---|
| 15(1) | 5.6954(3) | 5.6920(3) | 4.9408(3) | 89.895(7) |
| 25(1) | 5.6959(3) | 5.6926(3) | 4.9414(3) | 89.883(7) |
| 35(1) | 5.6959(3) | 5.6925(3) | 4.9412(3) | 89.870(7) |
| 45(1) | 5.6959(3) | 5.6925(3) | 4.9423(3) | 89.874(7) |
| 55(1) | 5.6964(3) | 5.6930(3) | 4.9423(3) | 89.860(6) |
| 65(1) | 5.6954(3) | 5.6920(3) | 4.9415(3) | 89.847(6) |
| 75(1) | 5.6955(3) | 5.6921(4) | 4.9424(3) | 89.833(6) |
| 85(1) | 5.6949(3) | 5.6915(4) | 4.9419(3) | 89.846(6) |
| 100(1) | 5.6955(3) | 5.6921(4) | 4.9424(3) | 89.844(6) |
| 115(1) | 5.6961(3) | 5.6927(4) | 4.9431(3) | 89.840(6) |
| 130(1) | 5.6968(3) | 5.69344) | 4.9436(3) | 89.848(6) |
| 145(1) | 5.6966(3) | 5.6932(4) | 4.9437(3) | 89.838(6) |
| 160(1) | 5.6968(3) | 5.6934(4) | 4.9441(3) | 89.843(6) |
| 180(1) | 5.6963(3) | 5.6929(4) | 4.9442(3) | 89.842(6) |
| 200(1) | 5.6974(3) | 5.6940(4) | 4.9452(3) | 89.832(6) |
| 220(1) | 5.6976(3) | 5.6942(4) | 4.9457(3) | 89.833(6) |
| 240(1) | 5.6984(3) | 5.6950(4) | 4.9460(3) | 89.821(6) |
| 260(1) | 5.6987(3) | 5.6953(4) | 4.9465(3) | 89.819(6) |
| 280(1) | 5.6986(3) | 5.6952(4) | 4.9463(4) | 89.811(6) |
| 290(1) | 5.6994(3) | 5.6960(4) | 4.9470(4) | 89.810(6) |
| 300(1) | 5.6997(3) | 5.6963(3) | 4.9475(3) | 89.808(7) |
| 323(1) | 5.7015(3) | 5.6982(3) | 4.9490(3) | 89.800(7) |
| 373(1) | 5.7032(3) | 5.6999(3) | 4.9501(3) | 89.801(7) |
| 423(1) | 5.7048(3) | 5.7014(3) | 4.9512(3) | 89.799(7) |
| 473(1) | 5.7063(3) | 5.7029(3) | 4.9530(3) | 89.797(6) |
| 523(1) | 5.7069(3) | 5.7035(3) | 4.9539(3) | 89.791(6) |

**LeBail Refinements**

| T(K) | $R_p$ | $R_{wp}$ | $R_{exp}$ | $\chi^2$ | $R_{Bragg}$ |
|---|---|---|---|---|---|
| 15-523 | 10.8-14.7 | 16.0-21.9 | 12.13-12.22 | 1.72-3.24 | 0.07-0.31 |

**Table S7:** Unit-cell parameters evolution with temperature from 298-1123 K.

| T(K) | a(Å) | b(Å) | c(Å) | β(º) |
|---|---|---|---|---|
| 298(1) | 4.6861(3) | 5.6903(3) | 4.9408(3) | 89.964(8) |
| 323(1) | 4.6872(3) | 5.6907(3) | 4.9418(3) | 89.964(7) |
| 373(1) | 4.6889(3) | 5.6925(3) | 4.9433(3) | 89.957(7) |
| 423(1) | 4.6898(3) | 5.6942(3) | 4.9444(3) | 89.958(7) |
| 473(1) | 4.6914(3) | 5.6958(3) | 4.9455(3) | 89.955(7) |
| 523(1) | 4.6933(3) | 5.6973(3) | 4.9473(3) | 89.954(6) |
| 573(1) | 4.6948(3) | 5.6979(3) | 4.9482(3) | 89.947(6) |
| 623(1) | 4.6958(3) | 5.6994(3) | 4.9494(3) | 89.937(6) |
| 673(1) | 4.6969(3) | 5.7011(3) | 4.9506(3) | 89.933(6) |
| 723(1) | 4.6976(3) | 5.7021(4) | 4.9521(3) | 89.928(7) |
| 773(1) | 4.6988(3) | 5.7033(4) | 4.9531(3) | 89.926(7) |
| 823(1) | 4.7004(4) | 5.7048(4) | 4.9551(3) | 89.840(7) |
| 873(1) | 4.7043(4) | 5.7070(4) | 4.9584(3) | 89.773(6) |
| 923(1) | 4.7070(4) | 5.7085(4) | 4.9598(4) | 89.749(6) |
| 973(1) | 4.7112(4) | 5.7109(4) | 4.9629(4) | 89.701(6) |
| 1023(1) | 4.7148(4) | 5.7122(4) | 4.9657(4) | 89.681(6) |
| 1073(1) | 4.7203(4) | 5.7156(4) | 4.9707(4) | 89.629(5) |
| 1123(1) | 4.7235(4) | 5.7178(5) | 4.9725(4) | 89.600(5) |

**LeBail Refinements**

| T(K) | $R_p$ | $R_{wp}$ | $R_{exp}$ | $\chi^2$ | $R_{Bragg}$ |
|---|---|---|---|---|---|
| 298-1123 | 11.8-13.4 | 15.9-18.4 | 10.44-10.54 | 2.28-3.12 | 0.08-0.30 |

**$FeWO_4$ Synthesized Sample – Neutrons**

**Table S8:** Unit-cell parameters evolution with temperature.

| T(K) | a(Å) | b(Å) | c(Å) | β(º) |
|---|---|---|---|---|
| 1.97(3) | 4.68883(9) | 5.69197(14) | 4.94270(12) | 90.180(3) |
| 9.85(5) | 4.68878(9) | 5.69189(14) | 4.94266(12) | 90.179(3) |
| 17.74(5) | 4.68881(9) | 5.69184(14) | 4.94271(12) | 90.177(3) |
| 29.69(5) | 4.68876(9) | 5.69185(14) | 4.94288(12) | 90.175(3) |
| 39.75(5) | 4.68856(9) | 5.69199(14) | 4.94308(11) | 90.172(3) |
| 49.87(5) | 4.68854(9) | 5.69215(14) | 4.94337(11) | 90.169(3) |
| 59.88(5) | 4.68833(9) | 5.69244(14) | 4.94377(11) | 90.164(3) |
| 69.79(11) | 4.68843(9) | 5.69254(13) | 4.94408(11) | 90.159(3) |
| 79.84(6) | 4.68841(9) | 5.69290(13) | 4.94432(11) | 90.158(3) |
| 89.66(16) | 4.68859(9) | 5.69280(14) | 4.94450(12) | 90.153(3) |
| 99.30(40) | 4.68884(9) | 5.69296(14) | 4.94478(11) | 90.154(3) |
| 118.50(90) | 4.68901(9) | 5.69328(14) | 4.94508(14) | 90.145(3) |
| 138.1(1.1) | 4.68955(9) | 5.69337(14) | 4.94552(11) | 90.143(3) |
| 157.9(1.1) | 4.69005(9) | 5.69368(14) | 4.94608(11) | 90.141(3) |
| 177.8(1.2) | 4.69060(9) | 5.69392(14) | 4.94659(11) | 90.133(3) |
| 198.3(7) | 4.69116(8) | 5.69445(13) | 4.94743(11) | 90.127(3) |
| 217.8(1.2) | 4.69172(8) | 5.69464(13) | 4.94787(11) | 90.122(3) |
| 237.7(1.2) | 4.69235(9) | 5.69518(14) | 4.94839(11) | 90.125(3) |
| 257.6(1.1) | 4.69303(9) | 5.69551(14) | 4.94916(11) | 90.122(4) |
| 277.4(1.1) | 4.69375(9) | 5.69617(14) | 4.94980(11) | 90.122(4) |
| 297.2(1.0) | 4.69434(9) | 5.69659(13) | 4.95052(11) | 90.118(4) |
| 316.9(1.0) | 4.69505(9) | 5.69737(14) | 4.95115(11) | 90.126(4) |
| 337(4) | 4.69572(9) | 5.69812(14) | 4.95189(12) | 90.128(4) |
| 370(4) | 4.69694(10) | 5.69892(14) | 4.95300(12) | 90.136(4) |
| 404(5) | 4.69800(11) | 5.69999(14) | 4.95430(14) | 90.143(4) |
| 438(5) | 4.69925(10) | 5.70124(15) | 4.95576(15) | 90.159(3) |
| 470(8) | 4.70048(11) | 5.70237(15) | 4.95697(15) | 90.172(3) |

**Rietveld Refinements**

| T(K) | $R_p$ | $R_{wp}$ | $R_{exp}$ | $\chi^2$ | $R_{Bragg}$ |
|---|---|---|---|---|---|
| 2-470 | 5.74-8.69 | 8.12-11.9 | 2.01-4.08 | 8.04-18.1 | 17.8-21.5 |